\documentclass[12pt,preprint]{aastex}

\slugcomment{To appear in the Astrophysical Journal.}

\shorttitle{Interstellar $^7$Li/$^6$Li Ratios}
\shortauthors{Knauth, Federman, \& Lambert}

\begin{document}


\title{An Ultra-High-Resolution Survey of the Interstellar $^7$Li-to-$^6$Li
Isotope Ratio in the Solar Neighborhood} 


\author{David C. Knauth\altaffilmark{1}$^,$\altaffilmark{2}$^,
$\altaffilmark{3}, S. R. Federman\altaffilmark{1}$^,$\altaffilmark{3},
and David L. Lambert\altaffilmark{4}}

\altaffiltext{1}{Department of Physics and Astronomy, University of Toledo, 
2801 W. Bancroft, Toledo, OH 43606; sfederm@uoft02.utoledo.edu.}
\altaffiltext{2}{Current address: {\it FUSE} Science Center, Department of 
Physics and Astronomy, The Johns Hopkins University, 3400 N. Charles St. 
Baltimore, MD 21218; dknauth@pha.jhu.edu.}
\altaffiltext{3}{Guest Observer, McDonald Observatory, University of Texas at 
Austin.}
\altaffiltext{4}{Department of Astronomy, University of Texas at Austin, 
Austin, TX 78712; dll@astro.as.utexas.edu.}

{}
\vfill

\begin{abstract}
In an effort to probe the extent of variations in the interstellar 
$^7$Li/$^6$Li ratio seen previously, ultra-high-resolution (R $\sim$ 360,000),
high signal-to-noise spectra of stars in the Perseus OB2 and Scorpius OB2 
Associations were obtained.  These measurements confirm our earlier findings
of an interstellar $^7$Li/$^6$Li ratio of about 2 toward $o$~Per, the value 
predicted from models of Galactic cosmic ray spallation reactions.  
Observations of other nearby stars yield limits consistent with the isotopic 
ratio $\sim$ 12 seen in carbonaceous chondrite meteorites.  If this ratio 
originally represented the gas toward 
$o$~Per,  then to decrease the original isotope ratio to its current value
an order of magnitude increase in the Li abundance is expected, but is not
seen.  The elemental K/Li ratio is not unusual, although Li and K are formed 
via different nucleosynthetic pathways.  Several proposals
to account for the low $^7$Li/$^6$Li ratio were considered,
but none seems satisfactory.

Analysis of the Li and K abundances from our survey highlighted two sight 
lines where depletion effects are prevalent.  There is evidence for enhanced 
depletion toward X~Per, since both abundances are lower by a factor of 4 when
compared to other sight lines.  Moreover, a smaller Li/H abundance is observed
toward 20~Aql, but the K/H abundance is normal, suggesting enhanced Li 
depletion (relative to K) in this direction.  Our results suggest that the 
$^7$Li/$^6$Li ratio has not changed significantly during the last 4.5 billion
years and that a ratio $\sim$ 12 represents most gas in the solar 
neighborhood.  In addition, there appears to be a constant
stellar contribution of $^7$Li, indicating that one or two 
processes dominate its production in the Galaxy. 
 
\end{abstract}


\keywords{ISM: atoms -- ISM: abundances -- Galaxy: open clusters and associations: 
individual (Per~OB2, Sco~OB2) -- Galaxy: solar neighborhood --
Stars: individual ($o$~Per, X~Per, $\chi$~Oph, $\zeta$~Oph, and 20~Aquilae)}


\section{Introduction}

One of the primary goals in astronomy centers on the origin of the elements, 
but a complete picture remains elusive.  Our focus is an improved understanding
of light element synthesis through observations of interstellar lithium in diffuse clouds. 
Since models of Big Bang nucleosynthesis (BBN) yield 10\% of the current
abundance of $^7$Li and negligible amounts of $^6$Li (Suzuki, Yoshi, \& Beers
2000), much theoretical effort has gone into finding the source for Li.
In the 1970's, the work of Reeves and collaborators (Reeves,
Fowler, \& Hoyle 1970; Meneguzzi, Audouze, \& Reeves 1971, hereafter (MAR);
Reeves et al. 1973; Reeves 1974) provided an alternate method of light element
production which plays an important role throughout the history of the Galaxy.
This process is now known as standard Galactic cosmic ray (GCR) spallation.
Light elements are created when interstellar C, N, and O nuclei are broken
apart by Galactic cosmic rays, the spallation process, and by 
$\alpha$~$-$~$\alpha$ fusion reactions.  The exact nature of the cosmic ray
source is still under debate (e.g., Ramaty et al. 2000a; Fields et al. 2001).
Two commonly accepted creation mechanisms for cosmic rays involve supernovae; p
and $\alpha$ particles are produced in and then accelerated by the explosion or
are sputtered off interstellar grains and then accelerated.

A low energy cosmic ray (LECR) component, accelerated C and O nuclei, may also
be present (Ramaty, Kozlovsky, \& Lingenfelter 1996).  Lithium, beryllium, and 
boron are produced through the inflight
fragmentation of C and O nuclei during collisions with ambient interstellar H
and He (Cass\'{e}, Lehoucq, Vangioni-Flam 1995; Higdon, Lingenfelter, \& Ramaty
1998; Lemoine, Vangioni-Flam, \& Cass\'{e} 1998; Lingenfelter, Ramaty, \& 
Kozlovsky 1998; Vangion-Flam et al. 1998; Parizot \& Drury 1999; Ramaty \& 
Lingenfelter 1999; Vangioni-Flam, Cass\'{e}, \& Audouze 2000).  The inverse of 
standard GCR reactions, this LECR component originates from core collapse 
supernovae (SN~{\small II}) grouped in a superbubble (Higdon et al. 1998; 
Lingenfelter et al. 1998).  Both standard GCR and LECR/superbubble models are 
able to account for the present day abundance of $^6$Li, $^9$Be, and $^{10}$B, 
but only 10 $-$ 25\% of $^7$Li and about 50\% of $^{11}$B (e.g. MAR; Ramaty et 
al. 1997).

The rapid rise in the $^7$Li abundance for stars with [Fe/H]~$\geq$~-0.5 
indicates the existence of a stellar source of $^7$Li.  Proposed 
stellar sources include asymptotic giant branch (AGB; Smith \& Lambert 1989, 
1990; Plez, Smith, \& Lambert 1993) and red giant branch (RGB; Smith et al. 
1995) stars, with contributions from SN~{\small II} and possibly novae.  In AGB
and RGB stars rapid transport of $^7$Be (Cameron \& Fowler 1971; Boothroyd, 
Sackmann, \& Wasserburg 1994, 1995; Wasserburg, Boothroyd, Sackmann 1995) 
between the He~burning~shell and the envelope yields $^7$Li.
In Type II supernovae (SN II), the
flux of neutrinos becomes so great that the spallation of heavy nuclei, C, N,
and O, into light elements can occur.  Through spallation in the He and C 
shells, SN II can produce observable amounts of both $^7$Li and $^{11}$B (Woosley et al.
1990; Woosley \& Weaver 1995).  
Another possible thermonuclear source of $^7$Li is novae.  Novae are 
predicted to play a significant part in the $^7$Li abundance (Starrfield et al.
1978; Romano et al. 2001).
The nature of the processes involved in the 
production of $^7$Li is still unclear.  A goal of our survey of the 
interstellar $^7$Li/$^6$Li
ratio is to shed light on the importance of the various processes.
  
Interstellar Li was first detected in the early 1970's (Traub \& Carleton 
1973).  Since then there have been relatively few reported observations of 
interstellar Li~{\small I}.   The early detections (Traub \& Carleton 1973; 
Vanden Bout et al. 1978;  Snell \& Vanden Bout 1981; Hobbs 1984; White 1986) 
showed weak interstellar Li~{\small I} absorption, barely discernible from the
noise, with equivalent widths ($W_{\lambda}$s) on the order of at most a few 
m$\mbox{\AA}$. These early observations showed that the interstellar Li~{\small
I} abundance was similar along different lines of sight.  However, large 
uncertainties remained due to low signal-to-noise ratios and uncertain
corrections for ionization and depletion onto grains.   With the advent of 
modern detectors and more sophisticated observing techniques, reliable 
observations of $^7$Li, and the much weaker $^6$Li line, were possible 
(Ferlet \& Dennefeld 1984; Lemoine et al. 1993; Lemoine, Ferlet, \& 
Vidal-Madjar 1995; Meyer, Hawkins, \& Wright 1993; Knauth et al. 2000,
Howarth et al. 2002).  

The Li isotope ratio is crucial for studies of Galactic chemical evolution
(Reeves 1993; Steigman 1993) since it is free from uncertain corrections 
(e.g., ionization and depletion).  Published $^7$Li/$^6$Li ratios in the local 
ISM indicate that the ratio may vary from the Solar System value of 12.3 
(Anders \& Grevesse 1989), but the results are far from conclusive.  Four
different isotope ratios have been published for the line of sight toward 
$\zeta$~Oph: $\geq$ 25 (Ferlet \& Dennefeld 1984) and 6.8$^{+ 1.4}_{- 1.7}$ 
for a single component fit to the data (Meyer et al. 1993), Lemoine et al.
(1995) obtained 1.4$^{+1.2}_{-0.5}$ ($\pm$ 0.6) and 8.6 $\pm$ 0.8 ($\pm$ 1.4) 
and Howarth et al. (2002) reported an average $^7$Li/$^6$Li ratio of 
13.2 $\pm$ 6.3 for their two component fit.  Lemoine et al. (1993) found an 
isotope ratio of 12.5$^{+4.3}_{-3.4}$ toward $\rho$~Oph, consistent with the 
Solar System value.  Knauth et al. (2000) derived isotope ratios for two 
velocity components toward $o$~Per of 1.7 $\pm$ 0.3 and 3.6 $\pm$ 0.6 and an 
isotope ratio of 10.6~$\pm$~2.9 for a single component toward $\zeta$~Per.  
The latter ratio is marginally consistent with the value of 5.5$^{+1.3}_{-1.1}$
reported by Meyer et al. (1993).  As discussed in \S 2, an accurate template 
for the velocity components along a given sight line appears to be required 
before consistent results emerge.

In our earlier study (Knauth et al. 2000), high-resolution (R $\sim$ 180,000)
spectra were obtained on the stars $o$ and $\zeta$ Per in the Perseus OB2
Association.  These stars were chosen for their proximity to IC~348, an
active star-forming region.  These data showed that the $^7$Li/$^6$Li ratio 
varies from the Solar System value toward $\zeta$~Per to a value of 
approximately 2 toward $o$~Per.  $o$~Per resides closer to IC~348 than does 
$\zeta$ Per and has 
an order of magnitude higher flux of cosmic rays (Federman, Weber, \& Lambert 
1996).  Knauth et al. (2000) suggested that the ratio of 2 was clear evidence 
for newly synthesized lithium toward $o$~Per resulting from GCR spallation 
reactions.  These data barely resolve the 2 velocity components separated 
by $\sim$ 3 km s$^{-1}$ toward $o$~Per.  To refine these findings and to probe
directions with interstellar clouds separated by $\sim$ 1 km s$^{-1}$, we
obtained ultra-high-resolution (UHR) observations toward other stars in the
Perseus OB2 and Scorpius OB2 Associations.  The stars in Sco OB2 lie close to an
active star-forming region approximately centered on $\rho$~Oph.  Furthermore, 
$\lambda$~Ori was observed because it resides in a region of 
active star formation.  In addition, an UHR survey of interstellar 
K~{\small I} (Welty \& Hobbs 2001) indicated that Li~{\small I} could possibly 
be detected toward $\lambda$~Ori.  20~Aql lies relatively far from active 
regions of star formation and serves as a useful comparison.  
  
The main goals of the present study were to determine reliable $^7$Li/$^6$Li 
isotope ratios in order to probe the extent of variations in its value and to 
help constrain the roots of Li production.  Since
$^6$Li is produced primarily through GCR spallation reactions, the GCR
contribution to $^7$Li can be accurately determined.  Assuming that the BBN
contribution is known from observations of halo stars, the remaining $^7$Li
comes from Population {\small I} stars.  These assumptions enabled interstellar 
constraints to be placed on the stellar production of $^7$Li. 
The organization of this paper is as follows.  In \S2 we describe the
observations and the data reduction. The profile syntheses for all data sets are
discussed and the results are tabulated in \S3.  The isotope 
ratios, Li~{\small I} and K~{\small I} column densities, elemental abundances,
depletion, and elemental K/Li ratios for each sight line are given 
in \S4.  In \S5 we discuss implications for future Li studies and thermal vs. 
turbulent broadening.  The results of the $^7$Li/$^6$Li isotope ratios are 
discussed in \S6 and \S7.  In \S8 constraints 
on stellar sources of $^7$Li are presented.  Finally, the results are 
summarized and suggestions for future work are contained in \S 9. 

\section{Observations and Data Reduction}
High signal-to-noise, high-resolution spectra of the Li~{\small I} 
doublet are needed, since coupled to the relatively weak absorption, the 
fine structure separation of the $^7$Li~{\small I} doublet 
(3p~$^2P_{3/2,~1/2}$ $-$ 2s $^2S_{1/2}$ at 6707.764 \AA\ for $J$ = 3/2 and at 
6707.915 \AA\ for $J$ = 1/2) is comparable to the isotope shift of $^6$Li 
($\sim$ 0.160 $\mbox{\AA}$). The result is a blend of the $^7$Li and $^6$Li lines 
(Ferlet \& Dennefeld 1984; Sansonetti et al. 1995).  The situation is further
complicated by the existence of several interstellar clouds along the line of 
sight (typically 6 clouds per kpc).  For this reason, UHR surveys [resolving
power (R) $\sim$ 600,000]  of K~{\small I} (Hobbs 1974a, 1974b; Welty \& Hobbs
2001) were used to find lines of sight with simple velocity structure (that is, 
one or two components) with significant amounts ($N$(K~{\small
I})~$\sim$~10$^{11}$ cm$^{-2}$) of absorption.  Assuming a constant K/Li ratio
(White 1986; Welty \& Hobbs 2001), these surveys yielded lines of sight to 
pursue in the quest for interstellar Li~{\small I} absorption.  
Table~\ref{stars-table} gives
the stellar data for the objects studied here. 

In order to extract the interstellar $^7$Li/$^6$Li ratio, knowledge about the 
velocity structure for a given line of sight (LOS) is necessary.  Data on 
another species are required for use as a template of the
velocity structure (Lemoine et al. 1993, 1995; Lambert et al. 1998; Knauth et 
al. 2000, Howarth et al. 2002).  We chose K~{\small I} because it 
has a similar ionization potential and is likely to reside in the
portion of the interstellar cloud containing Li~{\small I} (White 1986; Welty
\& Hobbs 2001).  The method used here differs from that of Lemoine et al. 
(1993, 1995) who adopted K~{\small I} $\lambda$7699 as their velocity 
template.  Instead, K~{\small I} $\lambda$4044 was chosen since it is
of comparable strength to the Li~{\small I} lines, although $\lambda$7699
is still useful in revealing weaker structure that may be visible within the 
$\lambda$4044 line profile.  Table~\ref{IPfw-table} lists
the ionization potential, IP, (Morton 1991), laboratory wavelengths for
Li~{\small I} and K~{\small I} (Morton 1991; Sansonetti et al. 1995), and the
oscillator strengths for Li~{\small I} and K~{\small I} fine structure lines
(Morton 1991) used in this study.  The oscillator strengths for the Li~{\small
I} hyperfine transitions were extracted from the fine structure oscillator
strengths (e.g., Welty, Hobbs, \& Kulkarni 1994).  The hyperfine structure of
K~{\small I} $\lambda$4044 was found to be negligible ($\sim$ 0.002 m$\mbox{\AA}$).

\subsection{Ultra-High-Resolution Observations}
\subsubsection{McDonald Observatory}
All observations were taken with the double-pass spectrograph of the 6-foot
camera  on the 2.7 m Harlan J. Smith Telescope at the University of Texas
McDonald Observatory.  Two spectral regions were observed, one centered on
K~{\small I} $\lambda$4044 and the other on Li~{\small I} $\lambda$6708.  The
former setting allowed a wavelength coverage of approximately 1.5~$\mbox{\AA}$,
while the latter yielded 2.5~\AA\ of spectrum.  Since such a narrow wavelength 
range was imaged, care was needed to ensure that interstellar features with 
large Doppler motions with respect to the Earth remained within the range of 
the detector.  For this purpose, hollow cathode lamps containing Th-Ar and Li,
as well as the solar spectrum, were utilized to determine the spectrograph 
settings.  Only a single weak, uncataloged line appeared in the Th-Ar spectrum
at $\lambda$6708.  Therefore, a Li hollow cathode was used for wavelength
calibration, even though the pressure broadened Li lines could not be used to 
determine the instrumental width.  An echelle grating was combined with a cross
disperser, which was used in 2$^{nd}$ order for the blue setting and in 
1$^{st}$ order for the red one.  A CuSO$_4$ (RG610) blocking filter for the 
blue (red) setting was placed behind the slit of the spectrograph to isolate 
the light from a specific order.  A slit width of 145~$\mu$m (slit \#2) was
used.  The echelle grating position was moved each 
night to minimize the effects of cosmic rays as well as any flat field 
artifacts, in hindsight a very wise decision.  Stellar exposures of 30$^m$ or 
less also minimized the confusion caused by cosmic rays.  For many of the 
observing runs, a stellar flat from a bright star was obtained.  Unfortunately,
due to poor weather conditions, a stellar flat was not obtained for every 
observing run.  This combination of procedures enabled the detection of 1\% CCD 
defects and minimized their effect on the data.   

The resolution was determined in the course of the analysis by the full width 
at half maximum (FWHM) of the Th line at 4043.395 $\mbox{\AA}$.  The FWHM was 
determined to be 0.01172~\AA\ with 7 pixels per resolution element at this 
wavelength setting.  While a lithium hollow cathode provided the wavelength 
calibration for the 6708 \AA\ region, the lithium lines are dominated by
pressure broadening (FWHM $\approx$ 0.0425 $\mbox{\AA}$).  Since the same 
spectrograph was utilized for both sets of observations and the resolution is
expected to be independent of the wavelength region, the resolution determined 
from the K~{\small I} setting was used for the Li~{\small I} setting.
At this resolution, the effects of thermal gas motions must be 
considered in determining an accurate instrumental width.  After correcting 
for thermal broadening at an assumed temperature of 300 K, the instrumental 
width was found to be 0.833 km s$^{-1}$, which corresponds to an R of 360,000.
This resolution was chosen to discern interstellar clouds separated by 
$\approx$ 1~km~s$^{-1}$.   

The reddened stars, $o$~Per, 40~Per, X~Per, $\zeta$~Oph, and $\chi$~Oph, in the 
Perseus OB2 and Scorpius OB2 Associations were observed, as were $\lambda$~Ori
and 20~Aql.  Data were also acquired on the bright unreddened stars
$\gamma$~Cas, $\alpha$~Leo, and $\alpha$~Vir.    All the observations were
obtained over eight observing runs between August 1998 and December 2000.
Table~\ref{obs-table} contains the following information for
each observing run: the stars and species observed, the observatory, the dates
of the observing run, the total exposure time for all observations, and the
signal-to-noise ratio (SNR) per pixel.  Also noted in Table~\ref{obs-table} are
the SNR per resolution element of the final summed spectra for all stars.
 
\subsubsection{Anglo-Australian Observatory}
Additional data on Li~{\small I} and the strong K~{\small I} line at 
7699 \AA\ were obtained (Lemoine 1995) from the Anglo-Australian 
Observatory (AAO) archive.  Li {\small I} and K~{\small I} were observed 
toward the stars, $\chi$ and $\zeta$~Oph, while only K~{\small I} was 
observed toward the stars, $o$~Sco, $\rho$~Oph, HD~154090, HD~165024, and 
$\mu$~Sgr.  A synopsis also appears in Table~\ref{obs-table}.

The K~{\small I} data were reduced to verify the accuracy of our Li~{\small I}
data reduction.  Comparison of the measured values of $W_{\lambda}$ for
K~{\small I} $\lambda$7699 with those presented in Welty \& Hobbs (2001),
based on the same data, show excellent agreement with the exception of the
absorption toward $o$~Sco and HD 154090, which differ by approximately 10\%.  
This overall agreement lends confidence that our lithium measurements toward
$\chi$ and $\zeta$~Oph are reliable. 

\subsection{Standard Data Reduction}

The data were reduced in a standard way utilizing the {\bf NOAO SUN/IRAF}
software (Revision 2.11.3).  Dark, bias, and flat lamp exposures were taken each
night to remove any instrumental effects due to the CCD detector.  Comparison 
spectra were taken periodically throughout the night, typically every two hours.
The average bias exposure was subtracted from all raw stellar, comparison
(Th-Ar), and flat images.  The scattered light was fit by a low order
polynomial, in both the dispersion direction and perpendicular to it for the
multi-order observations, and removed.  The pixel-to-pixel sensitivity was
taken into account by dividing the normalized flat into the stellar 
spectra.  The normalized flat is the average of 10 $-$ 20 flat lamp 
(5 $-$ 10 stellar flat) exposures, with each exposure having a flux level 
3 $-$ 5 times that of a single stellar exposure.  Next,
the pixels perpendicular to the dispersion were summed in each order for each
stellar and comparison lamp exposure.  The extracted spectra were placed on an
appropriate wavelength scale using the Th-Ar or Li comparison spectra, and
Doppler-corrected.  The spectra were then coadded and normalized to unity with 
a low order polynomial, yielding a
final spectrum with high signal to noise (based on the root-mean-square of the
deviations in the continuum), approximately 1000:1 per resolution element.  
Over the narrow wavelength range covered by these spectra, there were at most 
slight variations in the continua.

Each stellar spectrum was carefully examined for flat field artifacts or cosmic
rays before the final spectrum was created.  Cosmic rays that survived the 
reduction were removed.  When the cosmic rays coincided with the interstellar 
features in a spectrum, as was the case with 3 exposures of $\zeta$~Oph, 2 
exposures of 20~Aql and 1 exposure of $o$~Per, the affected spectra
were not included in the final sum.  The standard reduction process was
successful in extracting the data for most of the observing runs. While
examining the spectra for cosmic rays, the presence of two 1\% features (CCD
defects) that were not removed during the flat fielding process were detected. 
An improved flat fielding technique using stellar spectra was applied before 
the remaining steps in the extraction process.

The defects were seen in all spectra, but for $\sim$ 60\% of the observing runs
the defects were far ($\sim$ 200 pixels) from where the interstellar line(s)
appeared.  The defects were a problem for the McDonald observing runs of August
1998, December 1999, and December 2000.  The basic solution followed the 
prescription for removal of fixed pattern noise
in spectra acquired with the Goddard High-Resolution Spectrograph (Cardelli
\& Ebbets 1994).  The spectrum of an unreddened star was used as a template;
removal of the defect was accomplished by dividing the affected spectrum by the
template in pixel space, since defects remain fixed in pixel space.  Multiple 
observing runs of the interstellar species toward the same star gave an
additional check that proved useful in minimizing the effect of the 1\% defects.  
Where other measures exist for comparison (Meyer et al. 1993; Lemoine et al. 
1995; Knauth et al 2000; Howarth et al. 2002), there is good to excellent 
correspondence among spectra.  The McDonald spectra
were binned by three pixels to increase the SNR, a valid procedure since
there are 7 pixels per resolution element, and the AAO data for $\chi$
and $\zeta$~Oph were binned to the lower resolution of the McDonald data before 
combining them.

\section{Profile Synthesis}   
Each interstellar line profile yields the number of velocity components, the 
wavelength at line center, $\lambda$, FWHM, and $W_{\lambda}$, 
for each component.  The number of velocity components in the Li spectrum
was obtained from a 
visual examination of the K {\small I} spectrum.  For symmetric interstellar 
lines, the determination of the $W_{\lambda}$ is accomplished by fitting a 
Gaussian profile to the interstellar feature for comparison with the results
from the profile synthesis.  Asymmetric interstellar lines 
are telltale signs of multiple components along a given line of sight.  If more
than one component was present, the task of determining {\it W}$_{\lambda}$ 
involved simple Gaussian deblending.  Unresolved velocity structure is treated
as a single component and may result in slight differences in the derived 
quantities.  There are additional uncertainties in the calculated $V_{LSR}$ 
($\sim$ 0.5 km s$^{-1}$) derived from the K~{\small I} and Li~{\small I} lines.
This difference is attributed to the uncertainty in the wavelength calibration 
for the Li~{\small I} region, the result of there being larger widths for the 
pressure broadened, hollow cathode lines of Li~{\small I}.  For the stars, 
$\gamma$~Cas, $\alpha$~Leo, and $\alpha$~Vir, as well as 40~Per 
and $\lambda$~Ori, no detection of Li~{\small I} or K~{\small I} was evident.  
For these stars the 3-$\sigma$ upper limits, determined from the SNR and the 
FWHM of the instrument function, are listed in Table~\ref{UL-table} and will
not be discussed further.  Continuum placement uncertainties were not included 
in the error budget because variation in the continua was found to be 
insignificant.

If a single velocity component is present along 
a given sight line, then a simple Gaussian can be fitted to the line profile
in order to determine these three parameters.  A more realistic approximation 
of the profile function is given by a Voigt profile.  While a Voigt profile 
more precisely describes the wings of saturated absorption lines, Gaussian 
profiles were adequate in fitting the weak Li~{\small I} and K~{\small I} lines 
observed in this study.   If multiple components are
present, the determination of $b$-value,
$V_{LSR}$, and $N$ is more complicated. 

A C++ code (J. Zsarg\'{o} 2000, 
private communication) was utilized to compare the 
observed K~{\small I} and Li~{\small I} line profiles with synthetic ones.
The two profiles were synthesized independently.  Comparison of observed and
synthetic profiles was performed with a second-order gradient-expansion
Marquardt algorithm (Bevington \& Robinson 1992).
The synthetic profiles and $\chi^{2}$ were calculated with initial values of the 
parameters.  The gradient-expansion method
adjusts the initial values and the process is repeated with the updated values
until a change in $\chi^2$ of less than 0.01\% occurs. 
An additional feature of the program allowed any parameter to be fixed
in order to arrive at a solution.  One or more parameters were
fixed only when a high degree of certainty in the initial parameters was
present.   For details of the profile synthesis
see Zsarg\'{o} \& Federman (2002).  

Reliable initial 
values are crucial to finding the
most appropriate minima since there could be several minima caused by 
multiple velocity components or lower SNR.  The syntheses gave the $b$-value, 
$V_{LSR}$, and $N$ for each component as well as the total $W_{\lambda}$. 
Agreement between the observed $W_{\lambda}$ and that from the combination of
the components was at the 1-$\sigma$ level, with the exception of the 7 km 
s$^{-1}$ component toward $o$~Per (see \S3.2).  The
best fit was inferred from F-tests for trials based on
fits using different numbers of components. 
Figures~\ref{ofit1} $-$ \ref{2fit1} shows our UHR spectra obtained toward
$o$~Per, X~Per, $\chi$~Oph, $\zeta$~Oph, and 20~Aql, respectively.  From 
a visual examination of our spectra, only the clouds toward
$o$~Per reveal large $^6$Li abundances $-$ compare Figures~\ref{ofit1} $-$ 
\ref{2fit1}.  Profile synthesis provides a quantitative isotope ratio in all 
cases.    In addition, the enhanced $^6$Li abundance toward $o$~Per is 
clearly seen in our lower resolution spectrum (Knauth et al. 2000).

\subsection{$o$ Per}
\begin{figure}\figurenum{1}\epsscale{0.8}
\plotone{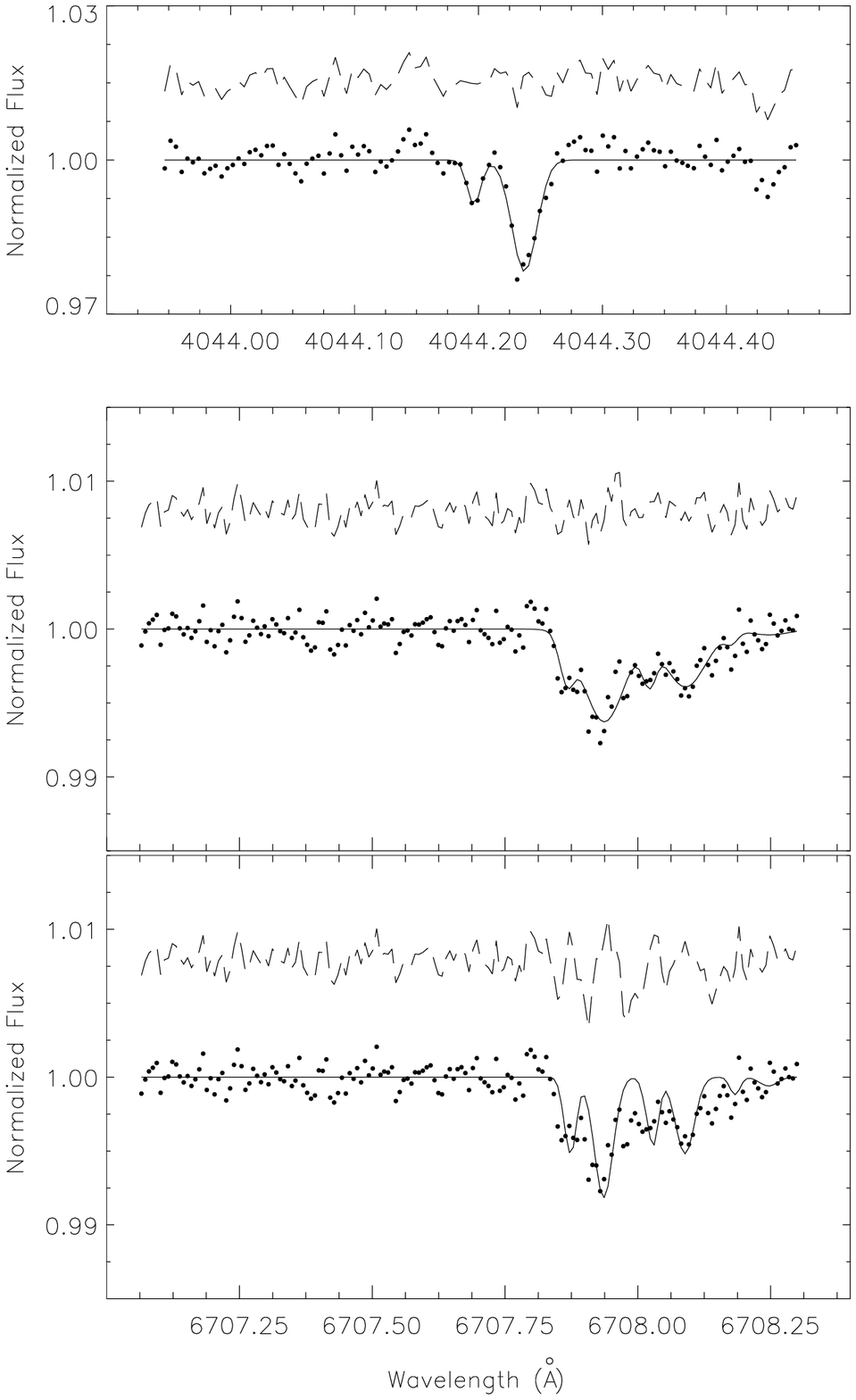}
\vspace{0.5in}
\caption{\label{ofit1}$Top$: the
preferred two component synthesis of the
K~{\small I} data toward $o$~Per using $b$-values determined from the line 
profile. $Middle$: the two component fit of Li~{\small I} using $b$-values 
determined from $^7$Li profile.  $Bottom$: velocity structure information from 
the K~{\small I} synthesis applied to Li~{\small I}.  In all plots, the data 
are represented by the filled circles, the fit is the solid line, and the dashed
lines are the residuals (data $-$ fit), offset to 1.02 and 1.08 for K~{\small I}
and Li~{\small I}, respectively.}
\end{figure}

\begin{figure}\figurenum{2}\epsscale{0.8}
\plotone{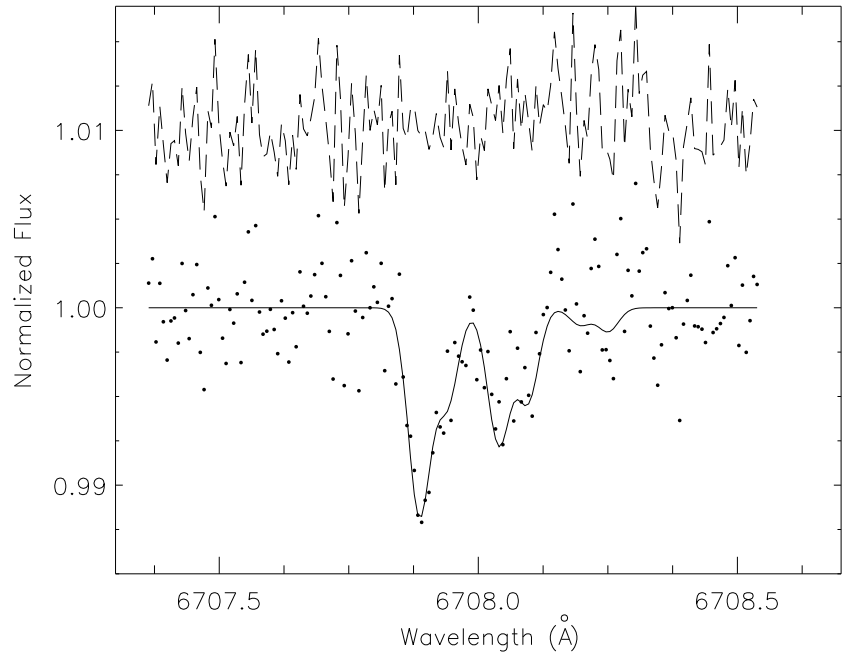}
\caption{\label{xper1}The synthesis of the 
Li~{\small I} data toward X~Per assumes two velocity components (S. R.
Federman 2001, private communication).  See Figure \ref{ofit1} for a 
description of the plot.  Residuals (data $-$ fit) are offset to 1.01 
(dashed line).}
\end{figure}

\begin{figure}\figurenum{3}\epsscale{0.8}
\plotone{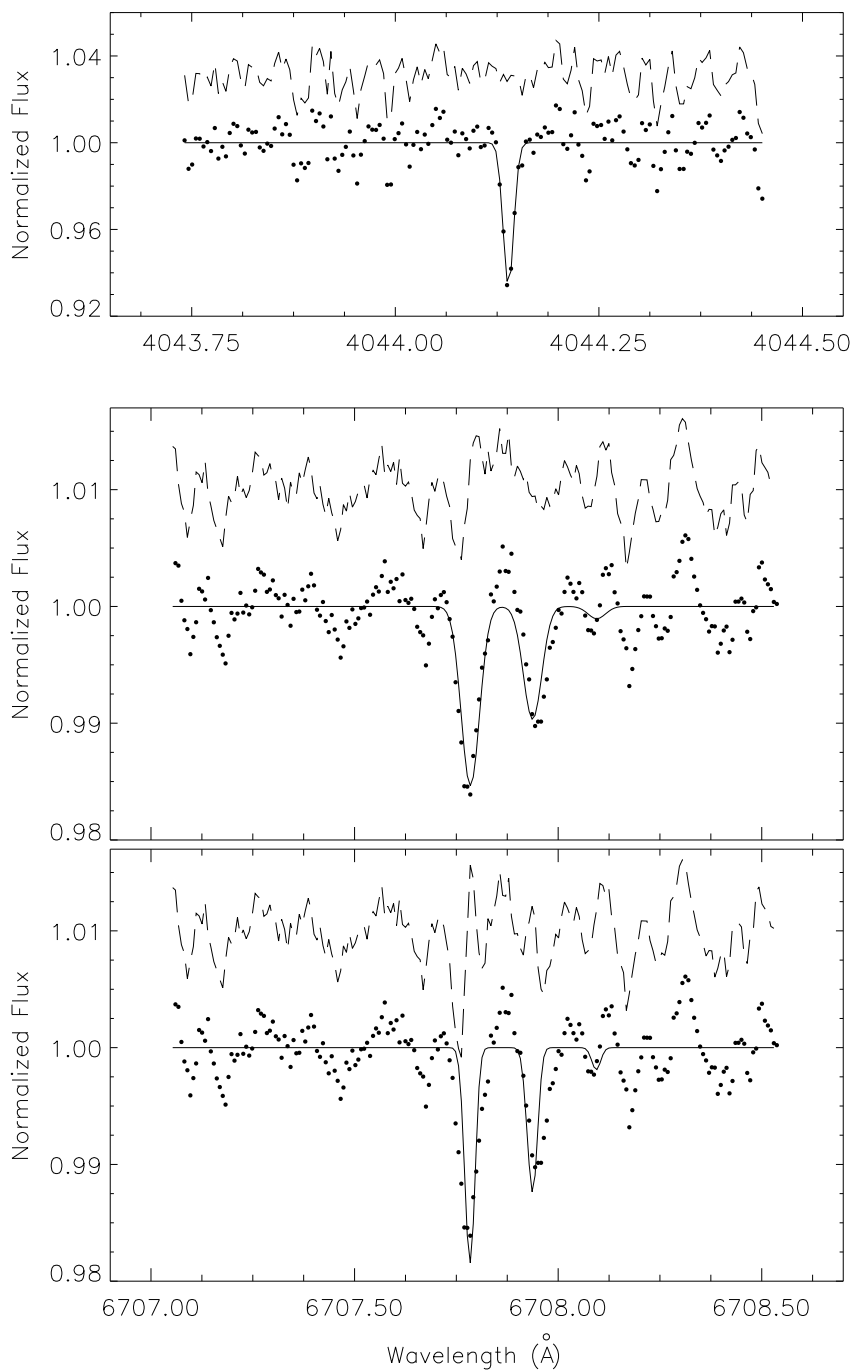}
\vspace{0.5in}
\caption{\label{coph1}$Top$: 
the one component synthesis of the K~{\small I} data toward $\chi$ Oph using 
$b$-values determined from the line profile.  $Middle$: one component 
fit of Li~{\small I} 
using $b$-values determined from $^7$Li profile.  The slight disagreement for 
the weaker $^7$Li line with the results of the synthesis is most likely caused 
by a CCD defect within the level of noise.   $Bottom$: fit of Li~{\small I} 
using velocity structure information from the K~{\small I} synthesis.  See 
Figure~\ref{ofit1} for a description of the plots. Residuals (dashed lines) for
K~{\small I} and Li~{\small I} are offset to 1.03 and 1.01, respectively.}
\end{figure}

\begin{figure}\figurenum{4}\epsscale{0.8}
\plotone{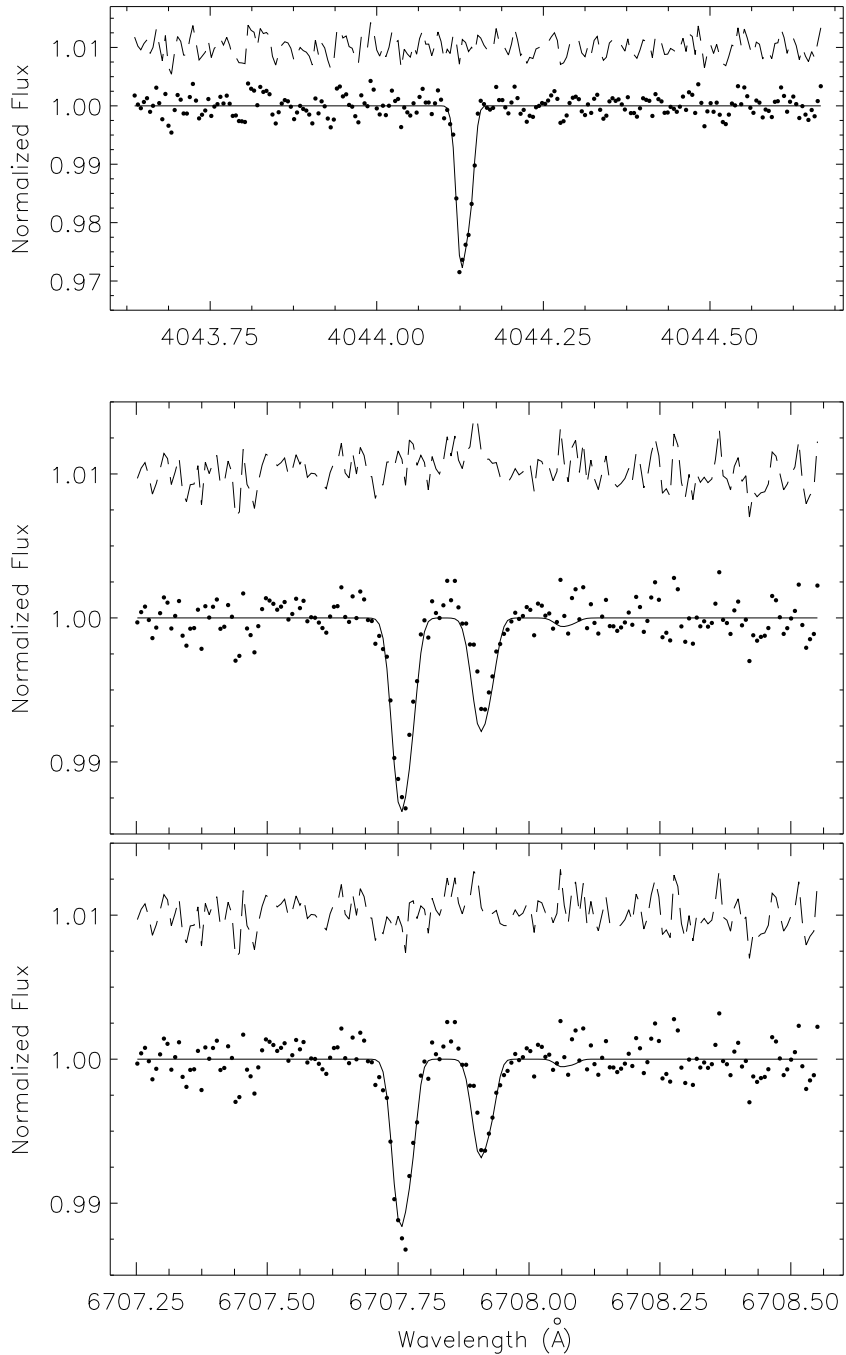}
\vspace{0.5in}
\caption{\label{zzoph1}$Top$:
the preferred two component synthesis of K~{\small I} using $b$-values from 
UHRS CN observations (Crawford et al. 1994) toward $\zeta$ Oph.  $Middle$: 
synthesis of Li~{\small I} using $b$-values determined from $^7$Li profile.  
$Bottom$: synthesis of Li~{\small I} using velocity structure information 
determined from K~{\small I}.  See Figure~\ref{ofit1} for a description of the 
plots.  Residuals (data $-$ fit) are offset to 1.01 (dashed lines) for both 
K~{\small I} and Li~{\small I}.}
\end{figure}

\begin{figure}\figurenum{5}\epsscale{0.8}
\plotone{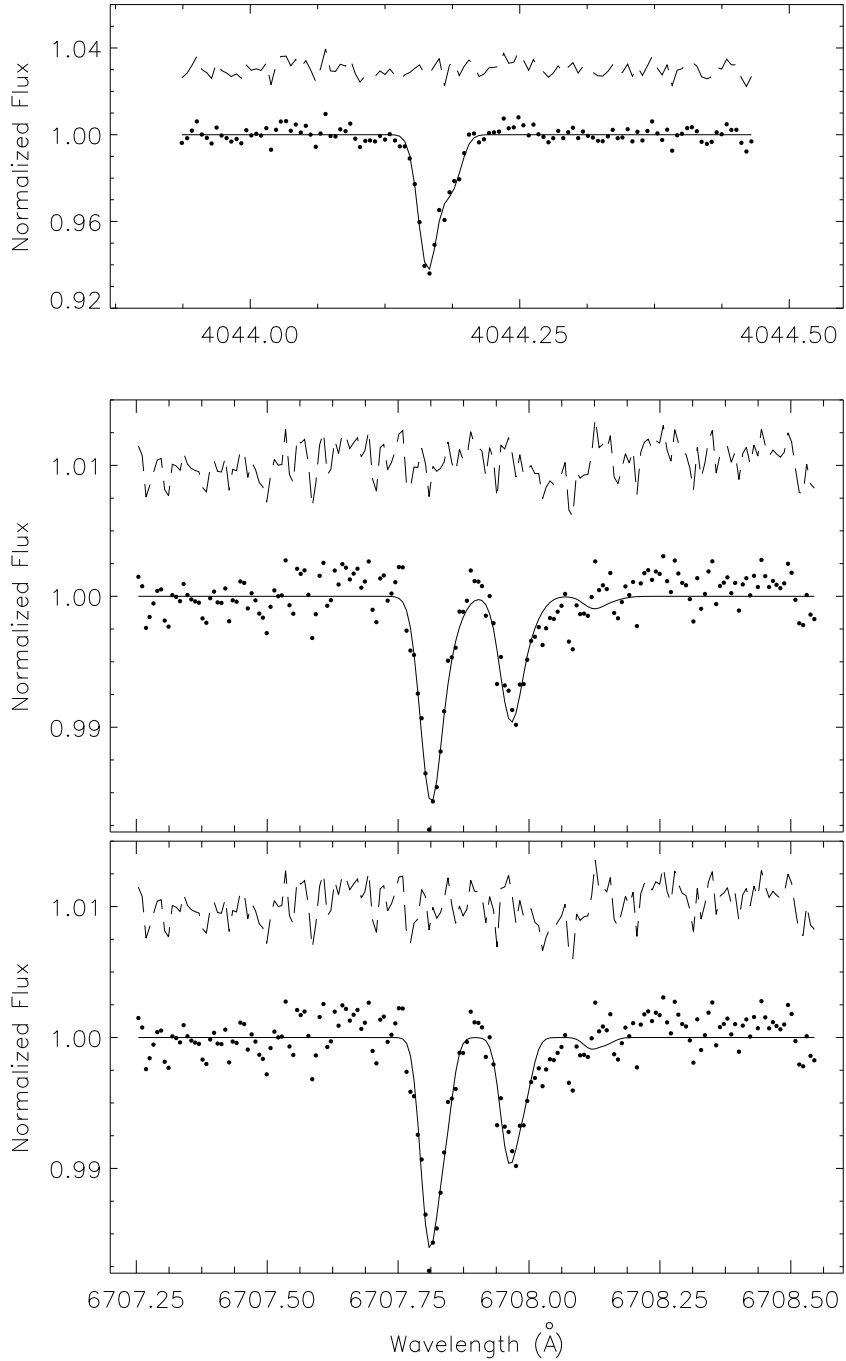}
\vspace{0.5in}
\caption{\label{2fit1}$Top$: the
two component synthesis of K~{\small I} toward 20~Aql using $b$-values 
determined from line profile. $Middle$: shows Li~{\small I} synthesis using 
$b$-values determined from $^7$Li profile.  $Bottom$: Li~{\small I} synthesis
determined from velocity structure information derived from the K~{\small I}.  
See Figure~\ref{ofit1} for a description of the plots.  Residuals (dashed 
lines) for K~{\small I} and Li~{\small I} are offset to 1.03 and 1.01, 
respectively.}
\end{figure}

\begin{figure}\figurenum{6}\epsscale{1.0}
\plotone{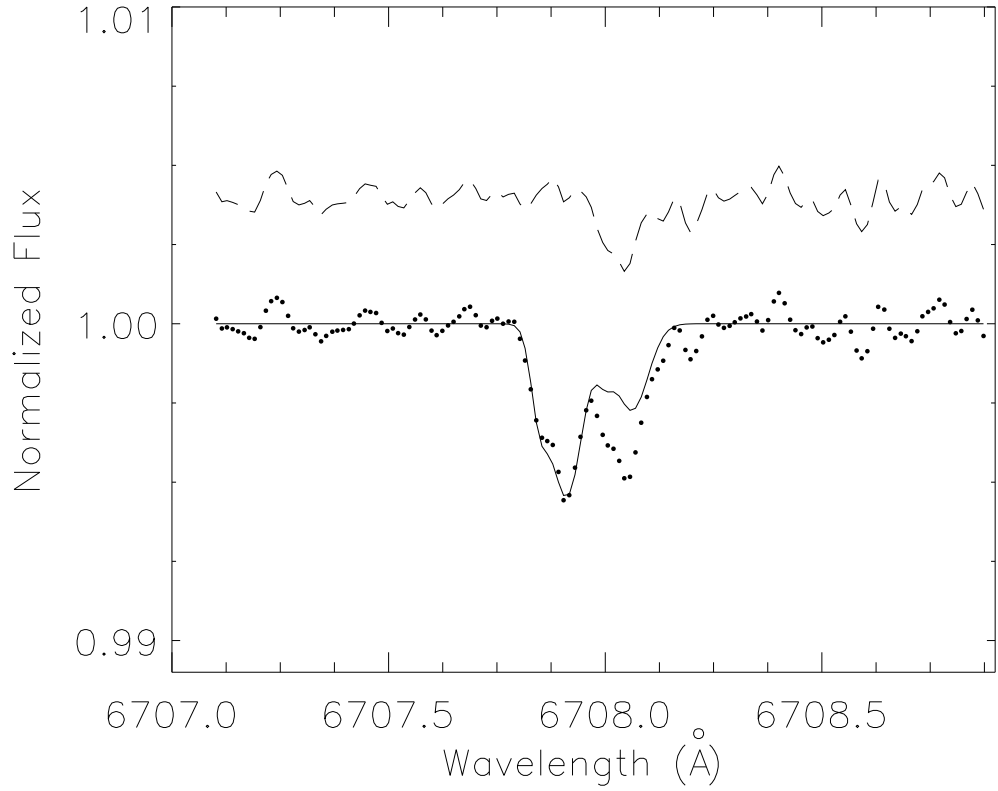}
\caption{\label{nature}We show our previous high-resolution Li~{\small I} 
spectrum (Knauth et al. 2000) with a $^7$Li/$^6$Li
ratio of 12, as a useful comparison.  It is quite clear that a
Solar System ratio provides a poor fit to the data toward $o$~Per.  See 
Figure~\ref{ofit1} for a description of the plot.  Residuals (data $-$ fit) 
are offset to 1.04 (dashed line).} 
\end{figure}

   At our resolution, a very weak absorption line provided by
   a single cloud of pure $^7$Li would be seen as a
   resolved doublet with the `blue' line twice the depth of the
   `red' line. Addition of $^6$Li increases the depth of the `red' line 
    which reduces the blue-to-red ratio of line depths below the
    2 to 1 ratio set by atomic physics (i.e., LS coupling). 
   Superficial inspection of the profiles at the telescope
   showed that all lines of sight but one were at or close
   to this 2 to 1 ratio. The exception was $o$~Per where the `red' line appeared 
   deeper than expected (Figure~\ref{ofit1}), a 
   telltale signature that either a high $^6$Li abundance in the cloud or
   the presence of a second higher velocity cloud contributing $^7$Li at
   about +7 km s$^{-1}$ from the main cloud could be masquerading as $^6$Li; 
   +7 km s$^{-1}$ is the isotopic shift. The weaker $^6$Li line, 
   which is about 0.16 \AA\ to the red of the weaker $^7$Li line,
   would be of value in determining the reason for the 
   deeper `red' line, but, unfortunately, the line
   is indistinguishable from the noise of the spectrum - see also
   the lower resolution spectra published by Knauth et al. (2000).  
   Figure~\ref{nature} shows our previous high-resolution Li~{\small I}
   spectrum (Knauth et al. 2000), fit with a synthetic spectrum assuming a 
   Solar System isotope ratio of 12 for both velocity components.   
   Examination of Figure~\ref{nature} clearly shows an enhanced $^6$Li 
   abundance.

   Our template K~{\small I} 4044 \AA\ profiles and those
   provided by observations of the K~{\small I} 7699 \AA\ line
   (Welty \& Hobbs 2001)
   serve to constrain the distribution
   of clouds toward $o$~Per. Our 4044 \AA\ spectrum
   shows two clouds separated by about 3 km s$^{-1}$ and
   both clouds are seen to contribute to the Li~{\small I}
   profile (Figure~\ref{ofit1}). Welty and Hobbs's higher resolution 
   observations of the 7699 \AA\ K~{\small I} line show that each of our
   clouds is a pair of clouds with a velocity separation
   of 1.0 km s$^{-1}$ for the weaker cloud and 1.3 km s$^{-1}$
   for the stronger cloud.  A fifth cloud, at a 
   velocity of about 2 km s$^{-1}$ to the red of the stronger
   line in Figure~\ref{ofit1}, has a column density of only 6\% of that in
   the stronger cloud. This fifth component is
   hidden by the noise in Figure~\ref{ofit1}. 

   Independent analyses of the Li~{\small I} and K~{\small I} profiles
   began with the introduction of the two velocity components clearly
   detected in Figure~\ref{ofit1}.  The reduced-$\chi^2$ ($\chi^2/\nu$)
   of both fits (Figure ~\ref{ofit1}) are at the 90 - 95\% confidence 
   level.  Derived parameters from these independent fits are given in 
   Table~\ref{fit_results}.  Considering the accuracy of the velocity scales, 
   the small difference in the cloud velocities derived from the K~{\small I} 
   and Li~{\small I} syntheses is not significant.  As can be seen in 
   Figure~\ref{nature}, a Solar System ratio for both velocity components
   yields a poor fit to the data.  A fit to the Li~{\small I} absorption
   was also made using the $V_{LSR}$ and $b$-values from the
   K~{\small I} synthesis (see bottom panel of Figure~\ref{ofit1}).
   Although the isotopic ratios are 
   altered slightly, the quality of this fit is poor
   ($\chi^2/\nu$ = 3.73) on account of the
   smaller $b$-values for K~{\small I}. This suggests (see below)
   that thermal broadening is the main contributor to the $b$-values.
   
   Our earlier spectra of the K~{\small I} and Li~{\small I}
   absorption were acquired at a factor of 2 lower resolution and 
   with a signal to noise ratio of 2500:1.  A two-cloud
   analysis gave isotopic ratios of 1.7 $\pm$ 0.3 and
   3.6 $\pm$ 0.6 for the 4 and 7 km s$^{-1}$ clouds,
   respectively (Knauth et al. 2000). Our result, for the 4 km s$^{-1}$ 
   component, of 2.1 $\pm$ 1.1 confirms the earlier result, but our current
   result of 8.1 $\pm$ 2.1 is in poor agreement with
   the earlier result.  The measured $W_{\lambda}$s for the blue component 
   of $^7$Li~{\small I} and both components of K~{\small I} agree well with our 
   previous measurements (Knauth et al. 2000).  However, the red component in Li
   is 0.1~m\AA\ larger than that obtained in Knauth et al. (2000).  The larger 
   $W_{\lambda}$ could be due to an undetected CCD defect.  Therefore, the 
   profile synthesis of the UHR data was also performed utilizing the results
   of Knauth et al. (2000).  This synthesis resulted in a reasonable fit to the
   data.  Although both syntheses are equally acceptable, the resulting column 
   densities only overlap at the 3-$\sigma$ level, suggesting that the 
   observational errors alone are inadequate in representing the true 
   uncertainty.  A Solar System isotope ratio for 
   both components is unlikely due to the almost one-to-one correspondence 
   between both members of the Li~{\small I} doublet (see 
   Figure~\ref{nature}).  
   
   As noted above, our K~{\small I} components are each a
   blend of two close components.  If these four clouds
   are included instead of two, constrained fits are possible using the 
   $b$-values from Welty \& Hobbs (2001), although the isotopic ratios are not 
   changed significantly.  Three component syntheses of 
   the K~{\small I} and Li~{\small I} line profiles resulted in an increased 
   $\chi^2$/$\nu$ (1.51 for K~{\small I} and 3.32 for Li~{\small I}) when 
   compared to the two component syntheses.  Application of an F test (Lupton 
   1993) shows that a third component is justified at the 70\% confidence level, 
   while a fourth component yields only a 50\% confidence.  In order to 
   investigate the possibility of $^7$Li masquerading as $^6$Li, another 
   synthesis was performed that incorporated the reddest component from Welty 
   \& Hobbs (2001).  This additional component was found to
   be indistinguishable from the noise in the 4044 \AA\ 
   line profile.  Again the quality of the fit is poor and
   does not change the occurrence of the low Li isotope ratio.  
  
   An additional probe of cloud structure along the line
   of sight is provided by the Na D profile
   (Welty \& Hobbs 2001), a more sensitive monitor of
   thin clouds than the 7699 \AA\ line. Four additional 
   clouds are seen in Na D: two to the red of the
   stronger K~{\small I} line in Figure~\ref{ofit1} and two to the
   blue of the weaker K~{\small I} line.  Na~{\small I} column densities in 
   these clouds do not exceed 1\% of the observed column 
   in the strongest of the five clouds detected in both 
   Na~D and K~{\small I}. Of relevance - perhaps - to the
   Li isotopic analysis is that the velocity shifts of
   additional clouds to the red are approximately
   equal to the velocity shift of $^6$Li from $^7$Li
   in the two main pairs of clouds. 
      
   These clouds of low Na~{\small I} column density appear
   at velocities relative to the main clouds that, if
   they contributed $^7$Li absorption, would reduce and even
   eliminate the need for $^6$Li absorption from
   the main clouds.  Another synthesis resulted in a reasonable fit to the data
   with limits to the $^7$Li/$^6$Li similar to the Solar System value for all
   velocity components.
   In order to achieve this, however, the ratio
   of $N$(Li~{\small I}) to $N$(Na~{\small I}) for the two reddest clouds
   must be about 10$^4$
   times that in typical clouds (Welty \& Hobbs 2001).
   Unfortunately, this argument replaces one
   astrophysical puzzle -- why is the $^7$Li/$^6$Li ratio
   so low?  -- for another -- why is the ratio of
   Li to Na atoms so high in the higher velocity
   clouds?  Inclusion of these low column density components 
   into the synthesis adds the additional complication that these 
   components contain little or no K~{\small I} absorption.
   It is interesting to note that the higher velocity
   clouds are seen in both Na~{\small I} and Ca~{\small II} toward $o$~Per, but
   only Ca~{\small II} absorption is detected toward $\zeta$ Per for which a 
   normal $^7$Li/$^6$Li ratio is found (Knauth et al. 2000). 
   The uncertainties and substitution of one puzzle for another suggest
   that the best results from the profile synthesis of the $o$~Per data are 
   given by the two component fit, indicating that the low relative strengths 
   for the $^7$Li doublet are the result of enhanced $^6$Li in one or both of 
   the clouds.

\subsection{Other Sight Lines}
The analyses of the other sight lines yielded lower limits to the
$^7$Li/$^6$Li ratio consistent with the Solar System value.  The weaker 
$^7$Li line was used to constrain the maximum
amount of $^6$Li present, since this feature
is a blend of the $^7$Li and $^6$Li lines and the ratio of $^7$Li is 2 to 1.
The profile synthesis produced a formal detection to $\chi$~Oph which is 
consistent with these limits.  The syntheses for these four sight lines also 
provided the means to extract the relative importance of thermal and turbulent
broadening from the $b$-values of Li and K.  We now describe the particulars 
for each sight line.

\noindent {\bf {\it X~Per}} $-$  At least one velocity component is clearly 
detected in Li~{\small I} absorption. Unfortunately, 
data on K~{\small I} $\lambda$4044 is not available for this sight line, a 
result of X~Per being the faintest star studied here.  Observations of 
K~{\small I} $\lambda$7699 reveal more complicated velocity structure (2
components with an R $\sim$ 200,000, S. R. Federman 2001, private communication;
4 components with an R $\sim$ 600,000, D. Welty 2001, private communication).
All three syntheses yield good fits, with $\chi^2/\nu$ at the 85 $-$ 95\%
confidence level.  An F test on the data revealed 95\% confidence in the 2 
component fit and 74\% confidence in the 4 component fit.  Higher S/N data are
necessary to distinguish the most appropriate velocity structure in 
Li~{\small I} absorption for this line of sight.  The best fit, with two 
velocity components, is displayed in Figure~\ref{xper1}; the results appear in 
Table~\ref{fit_results}.  The $b$-values,  1.20 km s$^{-1}$ and 1.15 km
s$^{-1}$, are similar to those derived from K~{\small I} $\lambda$7699 
absorption, 1.7 km s$^{-1}$ and 1.2 km s$^{-1}$ (S. R. Federman 2001, private 
communication), suggesting that turbulence dominates the interstellar 
clouds toward X~Per.  

\noindent {\bf {\it $\chi$~Oph}} $-$  Only one velocity component is present. 
Multiple velocity components are seen in UHR K~{\small I} $\lambda$7699 spectra
(Welty \& Hobbs 2001).  An unrealistic fit resulted (e.g., an emission 
line) from a two component synthesis of the K~{\small I} $\lambda$4044 
absorption.  A significantly weaker third velocity component is at least
2 km s$^{-1}$ away (Welty \& Hobbs 2001), but is within the noise level in our 
data.  The single component synthesis of the line profiles is shown in 
Figure~\ref{coph1}.  The slight disagreement for the weaker $^7$Li line with the
results of the synthesis is most likely caused by a CCD defect within the level 
of noise.  The $\chi^2/\nu$ for the top two fits are at the 90 $-$ 95\% 
confidence level.  The results of these independent syntheses are presented in 
Table~\ref{fit_results}.  The synthesis of Li~{\small I} with the 
smaller $b$-values derived from the K~{\small I} fit is not acceptable 
($\chi^2$/$\nu$ = 2.06); thermal broadening dominates this interstellar cloud.
As a test for a single primary component, a self-consistent $b$-value was 
calculated for K~{\small I} $\lambda$7699 utilizing $N$(K~{\small I}) derived
from $\lambda$4044 and the total $W_{\lambda}$ of the two velocity components 
of $\lambda$7699 (Welty \& Hobbs 2001).  This results in a $b$ = 0.75 km 
s$^{-1}$, which agrees with the $b$-value of Welty \& Hobbs (2001) within their 
quoted 0.1 km s$^{-1}$ uncertainty.

\noindent {\bf {\it $\zeta$~Oph}} $-$  Two velocity components, separated by 
$\sim$ 1~km~s$^{-1}$, are clearly
detected in both K~{\small I} and Li {\small I} absorption toward $\zeta$ Oph.
Figure~\ref{zzoph1} shows the results
of this two velocity component fit to K~{\small I} (top panel) and Li~{\small I}
(middle panel).  There is also a slight mismatch
between the fit and the data in the weaker $^7$Li line, which also occurred for 
$\chi$~Oph, most likely caused by a CCD defect within the level of noise.  The 
results of the two component fit are exceptional for K~{\small I} and quite 
good for the Li~{\small I} absorption, as shown in Table~\ref{fit_results}.
A single component with an equivalent width of 0.638 m\AA\ was reported toward
$\zeta$~Oph by Meyer et al. (1993), 0.55 m\AA\ for the combined two
components by Lemoine et al. (1995), and 0.68 m\AA\ for the combined two 
components of Howarth et al. (2002).  These measurements agree well with the
combined $W_{\lambda}$ for the two velocity components, 0.55 m$\mbox{\AA}$, 
reported here.  The K~{\small I} parameters fit the Li~{\small I} profile as 
well, suggesting that turbulent broadening dominates.  UHR data (Welty \& Hobbs
2001) show three additional relatively weak K~{\small I} $\lambda$7699 
absorption components [$N$(K~{\small I}) $\sim$ 1 $\times$ 10$^{10}$ cm$^{-2}$],
approximately 1~km~s$^{-1}$ to longer wavelengths and 1 km~s$^{-1}$ and 4 km 
s$^{-1}$ to shorter wavelengths than the two stronger components.  These 
relatively weak components are at the 
level of the noise in our $\lambda$4044 spectrum.  Inclusion of a third 
component into our K~{\small I} and Li~{\small I} profile syntheses, with the
parameters taken from Welty \& Hobbs (2001), results in only a 50\% confidence 
in there being another component.  Howarth et al. 
(2002) also find no evidence for a third component.

\noindent {\bf {\it 20 Aql}} $-$  There are two components clearly detected in 
K~{\small I} and $^7$Li~{\small I}.
Figure~\ref{2fit1} shows the fits to both K~{\small I} and Li~{\small I} using
each species to determine the profile parameters.  The results of the
independent fits are shown in Table~\ref{fit_results}.  The bottom graph
shows the fit to Li~{\small I} using the parameters derived from the K~{\small
I} synthesis.  This fit has a slightly larger reduced $\chi^2$ (1.36) because
the derived $b$-values from the Li~{\small I} fit are slightly
larger ($\sim$ 0.5 km s$^{-1}$) than the values found from the K~{\small I} fit,
indicating the importance of thermal broadening.  

\section{Analysis}

\subsection{$^7$Li/$^6$Li Ratios}  
The $^7$Li/$^6$Li ratios, a direct result of the
profile synthesis, are listed in Table~\ref{fit_results}.
The uncertainties were determined from the weighted
average of both the $^7$Li and $^6$Li column densities (when present).   The 
lower limits are derived from the expected relative line strengths for the 
$^7$Li doublet (2:1).  Toward $o$~Per, one
velocity component is nearly identical to the value predicted by models of GCR 
spallation reactions ($\sim$~2), while the other component may be similar to 
the Solar System value of 12.3 (see Knauth et al. 2000) as are the isotope
ratios, or their limits, toward X~Per, $\chi$~Oph, $\zeta$~Oph, and 20~Aql.   

Other measures of the isotope ratio are available for comparison.  The 
determination of $^7$Li/$^6$Li = 5.5$^{+1.3}_{-1.1}$ toward $\zeta$~Per,
from spectra with R $\sim$ 200,000 and SNR of 2100,
(Meyer et al. 1993) is marginally consistent with our previous observations
taken at a similar resolution and with a SNR of 2500 which resulted
in a $^7$Li/$^6$Li ratio of 10.6 $\pm$ 2.9 (Knauth et al. 
2000).  Toward $\rho$~Oph, where a single velocity component is present,
a ratio of 12.5$^{+4.3}_{-3.4}$ was reported (Lemoine et al. 1993) from
spectra obtained at R $\sim$ 100,000 and SNR of 2700.  
A single component toward $\zeta$~Oph yields $^7$Li/$^6$Li= 
6.8$^{+ 1.4}_{- 1.7}$ from spectra with R $\sim$ 200,000 and SNR of 2300
(Meyer et al. 1993), but it is actually two velocity components separated by 
1 km s$^{-1}$ (Lambert, Sheffer, \& Crane 1990; Crawford et al. 1994; 
Crawford \& Barlow 1996; Welty \& Hobbs 2001; this work).  Lemoine et al. 
(1995) reported isotope ratios of 1.4$^{+0.8}_{-0.5}$~($\pm$~0.6) and 
8.6~$\pm$~0.8 ($\pm$~1.4) for two velocity components separated by $\sim$
5 km s$^{-1}$ from spectra taken at R of 100,000 and with a SNR of 7500.
The latter is the single component described by Meyer et al. (1993).
In a recent paper, Howarth et al. (2002) report an average $^7$Li/$^6$Li
ratio of 13.2 $\pm$ 6.3 for their two component fit toward
$\zeta$~Oph based on spectra taken at R $\sim$ 10$^6$ and a signal-to-noise
ratio of 1200:1. 

Lemoine et al.'s results for the weaker bluer component at $V_{helio}$ = -19 
km s$^{-1}$ with the low isotope ratio toward $\zeta$~Oph raises a number of 
issues.  First, the $^7$Li~{\small I} feature is but a 2 to 3-$\sigma$ 
detection, and similar positive excursions appear in their spectrum (their
Figure~6).  Second, this component has a K/Li ratio much smaller than that found
here for this sight line and others.  Another problem is that there is no 
molecular material associated with the velocity component having a low isotope
ratio.  For all other lines of sight in which Li~{\small I} has been detected, 
substantial molecular material is found at the velocity of Li~{\small I} 
$\lambda$6707.  The component may be the result of using the strong K~{\small I}
line at 7699 /AA/ as the velocity template because it detects weaker 
interstellar features than will be seen in Li~{\small I}.  The weak K~{\small I}
$\lambda$7699 component is at the limit of our signal to noise in $\lambda$4044.
No evidence for the weak velocity component in Li~{\small I} toward $\zeta$~Oph 
is suggested by our data or that of Howarth et al. (2002).   These facts seem to
indicate that a velocity component with a low isotope ratio is not present 
toward $\zeta$~Oph.  It is important to note that the SNR of Lemoine et al.'s 
data is a factor of 2 larger than the SNR of our data.

\subsection{Neutral Columns of Li and K}
\begin{figure}\figurenum{7}\epsscale{1.0}
\plotone{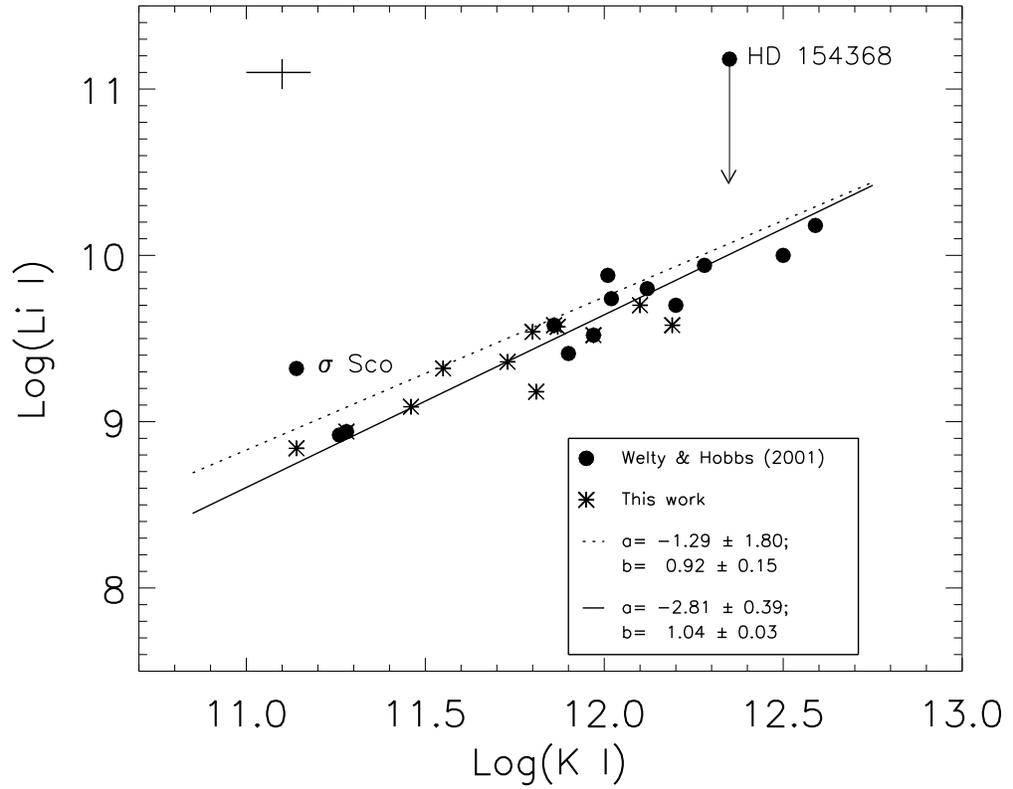}
\caption{\label{livsk}Plotted are the values of Log ($N$(K~{\small I})) vs Log
($N$(Li~{\small I})) for the lines of sight studied here (asterisks) and from 
Welty \& Hobbs 2001 (filled circles).  The dotted line is the 
best fit from Welty \& Hobbs (2001) and the solid line is the least 
squares fit to all the data.  A typical error bar is shown in the upper left 
of the graph.} 
\end{figure}

Figure~\ref{livsk} shows a comparison of $N$(Li~{\small I}) to
$N$(K~{\small I}).  The least squares fit (solid line) for all lines of sight
studied here and in Welty \& Hobbs (2001) is shown.  The dotted line is the
best fit from Welty \& Hobbs (2001).  Our least squares fit includes
the uncertainties from this work and Knauth et al. (2000), as well as the 
observational uncertainties for Li~{\small I} and an assumed 20\% uncertainty
in the K~{\small I} values (Welty \& Hobbs 2001 and references therein).  
Since a fit of data points derived from individual velocity structure 
(asterisks) has a higher correlation coefficient (r=0.92) than does the fit 
(r=0.81) based on total columns (solid circles), use of individual velocity 
components along the line of sight is the more appropriate means for comparison.
Another consideration is that some of the Li~{\small I} observations from the 
literature are only modest detections at best; higher quality data may also
improve the correlation coefficient.  Significantly greater dispersion about 
our fit arises when alternative component structures discussed in \S3 are 
included; we consider this further evidence for the preferred results given 
in Table~\ref{fit_results}.

The furthest data point from the fits is for the LOS toward HD~154368.  This 
LOS has the largest column of Li~{\small I} reported (Snow et al. 1996),
almost an order of magnitude larger than toward $\rho$~Oph, the next largest 
reported column density (White 1986; Lemoine et al. 1993; Knauth et al. 
unpublished). The measurement toward HD~154368 could be as much as a factor of 
5 too high (J. Black 2002, private communication), or K~{\small I} may not scale
linearly with Li~{\small I}. 

\subsection{Elemental Abundances}
The derivation of the total interstellar abundance requires knowledge of the
abundance of Li~{\small II} and the amount of depletion onto grains.  In
interstellar space, atoms of Li are predominantly singly ionized 
in view of the fact that the ionization potential of Li {\small I} is relatively
low (5.31 eV).  The ionization potential of Li~{\small II} is 75.6 eV.  
We consider the gas phase abundance here and wait to discuss 
implications for depletion onto grains (\S5.2).  An adequate estimate of the
Li abundance is obtained through ionization balance and 
an independent determination of the electron density ($n_e$).   The electron 
density is inferred directly from the column of C$^+$, the most abundant 
element providing electrons, the gas density ($n$) (Federman et al. 1994; 
Knauth et al. 2001), and the total proton column density 
[$N_{tot}$(H)~=~$N$(H~{\small I})~+~2$N$(H$_2$)] (Savage et al. 1977;
Bohlin et al. 1978; Diplas \&
Savage 1994) along the LOS.  Table~\ref{proton} contains the information on 
$N$(H~{\small I}), $N$(H$_2$), $N_{tot}$(H), and $n$.  Since no precise 
interstellar measurements of C$^+$ exist for the 
clouds toward $o$~Per, $\chi$~Oph, and 20~Aql, the weighted mean interstellar 
ratio of $N$(C$^+$)/$N_{tot}$(H) = (1.42 $\pm$ 0.13) $\times$ 
10$^{-4}$ (Sofia, Cardelli, \& Guerin 1997) was utilized.  Precise measures of 
$N$(C$^+$)/$N_{tot}$(H) are used in the analysis toward X~Per 
[(1.06 $\pm$ 0.38) $\times$ 10$^{-4}$; Sofia, Fitzpatrick, \& Meyer 1998] 
and toward $\zeta$~Oph [(1.32 $\pm$ 0.32) $\times$ 10$^{-4}$; 
Cardelli et al. 1993].  Welty \& Hobbs (2001) discussed the effects of large
molecules on ionization balance and found that depletion (or elemental 
abundances) of Li, Na, and K are not altered appreciably.

Through ionization balance we can determine the gas phase lithium abundance
(Li/H), without a correction for depletion onto grains, via

\begin{equation}
{A_g({\rm Li})} =  { {\rm Log} \left( \left[{N({\rm Li~I})} \over {N_{tot}({\rm H})}\right]   \left[{G({\rm Li~I})} \over {\alpha({\rm Li~II})n_e}\right] \right) + 12 }.
\end{equation} 

\noindent In this equation,  {\it A}$_g$(Li) is the elemental abundance,
$\alpha$(Li~{\small II}) is the radiative recombination rate constant
(P\'{e}quignot \& Aldrovandi 1986), {\it G}(Li~{\small I}) is the
photoionization rate from the ground state corrected for attenuation by dust 
grains.  The theoretical
determination of $\alpha$(Li~{\small II}) depends on the photoionization cross
section from all levels and is known to better than 2.5\%
(P\'{e}quignot \& Aldrovandi 1986).  The photoionization rates at the cloud
surface were determined from the measured ionization cross sections of Hudson \&
Carter (1965a, 1967a) and the average interstellar radiation field of Draine
(1978).  A similar expression holds for $A_g$(K), where K~{\small I} rates were 
determined from the measured ionization cross sections of Hudson \& Carter 
(1965b, 1967b), Marr \& Creek (1968), and Sandner et al. (1981).

The photoionization rates need to be corrected for the effect of attenuation
from dust in the interstellar cloud(s).  This is accomplished through use of
extinction curves for the particular line of sight.  If one were to assume that all dust particles are 
the same throughout the Galaxy, one would get an average Galactic 
extinction curve (Code et al. 1976).

From the extinction curves (Papaj, Krewlowski, \& Wegner 1991), the value of
selective extinction, $A_{\lambda}$, for the reddened star, can be derived
with the equation,

\begin{equation}
A_{\lambda*} = {{E(\lambda - V)} \over {E(B - V)}}  E(B - V) + A_{V*}.
\end{equation}

\noindent Eqn. 2 assumes there is no interstellar extinction along the line
of sight toward the unreddened star.  This is generally not the case; 
therefore we need to include terms for selective extinction for the 
unreddened star, $A_{\lambda_o}$ and $A_{V_o}$.  These values of selective 
extinction were obtained from the average Galactic extinction curve 
(Code et al. 1976) scaled to the $E(B~-~V)$ of the unreddened
star, typically 0.02 magnitudes.

Correcting the photoionization rates for attenuation involves a multiplicative
factor of the form exp(-$\tau_{\lambda}$), where $\tau_{\lambda}$ =
$A_{\lambda}$/2.  The factor of 2 arises because radiation impinges from both 
sides of the slab representing a cloud.  An optical depth ($\tau_{1200}$) was
chosen for Li~{\small I}, since 1200~\AA\ lies in the middle of the range in
wavelength that leads to its photoionization.  Two optical depths were used for
K~{\small I}: $\tau_{1200}$ for the overlap region and $\tau_{2500}$ for
wavelengths longer than 2400 $\mbox{\AA}$, which do not contribute to the photoionization
of Li~{\small I}.  The total attenuated photoionization rate 
for potassium comprises the sum for the two intervals.  The long wavelength
interval contributes about 5\% to the total.  For most of the observed lines of
sight, the extinction curves are similar to the standard extinction curve (Code
et al. 1976), the exceptions being X~Per and $\chi$~Oph.  These anomalous
extinction curves arise because X~Per is a member of an x-ray binary system and
$\chi$~Oph is an emission line star.  The different extinction laws change 
the photorates by approximately 20 $-$ 30\% compared to the standard extinction 
curve; larger excursions result for the anomalous extinction
curves of X~Per ($\sim$ 70\%) and $\chi$~Oph ($\sim$ 40\%).

The derived interstellar lithium abundances, shown in Table~\ref{depletion}, are
quite similar to other recent determinations with the exception of X~Per and
20~Aql.  The LOS toward X~Per exhibits almost an order of magnitude
lower Li abundance than other sight lines, while the Li abundance toward 20~Aql
is about a factor-of-2 less.  Using the above analysis for consistency, 
previous measures are: 2.9 $\times$ 10$^{-10}$ toward $o$~Per (Knauth et al.
2000); 3.3 $-$ 3.6 $\times$ 10$^{-10}$ toward $\zeta$~Per (Meyer et al. 
1993; Knauth et al. 2000); 2.7 $\times$ 10$^{-10}$ toward  $\rho$~Oph 
(Lemoine et al. 1993); and 3.9 $-$ 4.9 $\times$ 10$^{-10}$ toward 
$\zeta$~Oph (Meyer et al. 1993; Lemoine et al. 1995).  These values are 
different from those presented in the referenced work because they
include our correction for the attenuation of the radiation field and use a 
different photoionization rate.
  
\subsection{Depletion onto Grains}
The determination of the elemental Li abundance requires an uncertain correction
for the amount of depletion onto grains.  From the gas phase Li/H abundance, the
amount of Li depletion onto grains was estimated.  The amount of depletion is 
measured with respect to the Solar System value from meteorites, 
Log((Li/H)$_*$/(Li/H)$_{\odot}$), where (Li/H)$_{\odot}$ = 
20.5 $\times$ 10$^{-10}$ (Anders \& Grevesse 1989).  The K depletion was derived
in a similar manner using the value of 
(K/H)$_{\odot}$~=~1.26~$\times$~10$^{-7}$ (Anders \& Grevesse 1989).  One
problem with the use of the Solar System values for depletion measures is
the assumption that the Solar System value represents undepleted gas in
the solar neighborhood.  There is evidence (Wielen, Fuchs, \& Dettbarn 1996;
Wielen \& Wilson 1997) that the Sun actually formed closer to the center of 
the Galaxy, by about 2 kpc.  If our Sun formed in a more metal-rich 
environment, then the comparison with the Solar System values is inappropriate.
Furthermore, the Solar System values represent the interstellar gas at the epoch
of formation, 4.6 Gyr ago.

The amounts of depletion, given by the depletion indices $D$(Li) and $D$(K), are 
listed in Table~\ref{depletion}.  The indices for Li and K are similar for 
most sight lines ($\sim$~-0.9 dex), with the exceptions of those toward X~Per
and 20~Aql which are discussed in more detail below.  Welty \& Hobbs (2001) also
reported similar amounts of depletion for the two species, but for a larger
number of sight lines.  Their depletion indices are $D$(K) $\sim$ $D$(Li) $\sim$
-0.6 dex.  The use of a constant index for Li and K just scales the 
gas phase abundance without revealing any new information; direct
comparison to the Solar System may be more appropriate in studying 
effects of depletion onto grains.

\subsection{Elemental K/Li Ratio}
Through the comparison of 
$N$(Li~{\small I}) and $N$(K~{\small I}) $-$ see \S4.2, Welty \& Hobbs (2001) 
suggest a constant elemental K/Li abundance ratio.  It is straightforward to 
calculate the elemental K/Li abundance ratio and compare it with the Solar 
System value on a component-by-component basis for our results.  Since the 
amount of depletion onto grains is similar for the two species (Welty \& Hobbs
2001) for most directions, its uncertainty in Eqn. 1 can be minimized by taking 
the ratio of {\it N}(K~I) to {\it N}(Li~I).  This ratio, which also eliminates
$n_e$ and $N_{tot}$(H), is represented by

\begin{equation}
{ \left[{A_g({\rm K})} \over {A_g({\rm Li})}\right] } = 
{ \left[{N({\rm K\ I})} \over {N({\rm Li\ I})}\right] } 
{ \left[{\alpha({\rm Li\ II})} \over {\alpha({\rm K\ II})}\right] } 
{ \left[{G({\rm K\ I})} \over {G({\rm Li\ I})}\right] },
\end{equation}

\noindent where the observed values were compared to the Solar System value 
for {\it A}$_g$(K)/{\it A}$_g$(Li) of 61.6 (Anders \& Grevesse 1989).  Using the 
theoretical ratio for  $\alpha$(K~{\small II})/$\alpha$(Li~{\small II}) = 0.58 
(P\'{e}quignot \& Aldrovandi 1986), the K/Li ratio then mainly depends on the 
photoionization rates for  K~{\small I} and Li~{\small I}. 

For the best fit velocity structure (\S4.1), the calculated elemental K/Li
abundance ratios are consistent with the Solar System value of 61.6 (Anders 
\& Grevesse 1989), considering their
mutual uncertainties $-$ see Table~\ref{K/Li}.  On the other hand, both 
significantly higher and lower abundance ratios result when other component
structures are considered.
We, therefore, suggest that the K/Li ratio can be used
to discriminate between syntheses based on different velocity structure.   For 
the LOS toward X~Per, both Li and K abundances are low by a
comparable amount and thus the K/Li ratio is unaffected.  This suggests that 
the amount of depletion onto grains
is substantially greater or that $n_e$ has been underestimated.  The
one exception is the LOS toward 20~Aql; both components yield elemental K/Li
abundance ratios that are about two times larger.  As an additional check for
this line of sight, our previous high-resolution data on K~{\small I}
$\lambda$7699 (Knauth et al. 2001) were reanalyzed with the velocity component
structure deduced from our UHR data.  This reanalysis yielded similar
$N$(K~{\small I}) and elemental K/Li abundance ratios to those obtained from
$\lambda$4044.  Therefore the disparate K/Li ratios toward 20~Aql appear
to be real.  The discrepancy could be the result of an increase (decrease) in
the photoionization rate of Li (K), of evidence for enhanced depletion of Li
compared to K, or of the possibility that Li production (destruction) was lower
(higher) along this LOS.  The low K/Li ratios toward 20~Aql are discussed 
further in \S6.

\section{Interstellar Matters}
\subsection{Detecting Lithium}
At our resolution, velocity structure separated by $\sim$ 1 km s$^{-1}$ can be 
resolved (e.g. $\zeta$~Oph).  Most of the additional velocity structure 
detected in K~{\small I} $\lambda$7699 (Welty \& Hobbs 2001) is separated by 
more than 1~km~s$^{-1}$ and should be resolved in our data.  This situation is reminiscent
of the component structure utilized by Lemoine et al. (1995) for the gas 
toward $\zeta$~Oph.  Because its $f$-value is much larger,
K~{\small I} $\lambda$7699 reveals components too weak to see in absorption from
either K~{\small I} $\lambda$4044 or Li~{\small I} $\lambda$6707.  Although 
$\lambda$7699 is not an ideal template for velocity structure, it plays an
important role in the search for Li~{\small I} by revealing sight lines where
Li~{\small I} may be detectable.   

In addition to absorption from K~{\small I}, another useful means
for determining lines of sight having detectable amounts of
Li~{\small I} is the amount of molecular material present $-$ i.e., CH, C$_2$,
or CN.  Atoms with low IPs and molecules exist only in the relatively denser
regions of diffuse gas.  The molecular column densities for the 
lines of sight studied here are presented in Table~\ref{CHCNC2}.  For sight 
lines where Li~{\small I} is found, significant amounts of molecular 
absorption are also observed (e.g., $N$(CH) $\approx$ $N$(C$_2$) $\approx$
10$^{13}$ cm$^{-2}$; $N$(CN) $\geq$ 10$^{12}$ cm$^{-2}$).  Since 40~Per and
$\lambda$~Ori are in regions of active star formation and have relatively
strong K~{\small I} $\lambda$7699 absorption, they were considered potential
targets.  However, no Li~{\small I} was detected; of all stars observed in 
our survey, 40~Per has the lowest values for $N$(C$_2$) and $N$(CN) and 
$\lambda$~Ori has the lowest value of $N$(H$_2$) $-$ see Table~\ref{proton}.
Therefore the amount of molecular material present along a given LOS seems a
superior indicator for detectable amounts of Li~{\small I}.

\subsection{Thermal vs. Turbulent Broadening}
Atoms of K~{\small I} and Li~{\small I} seem to reside in the same portion of
interstellar clouds.  Through measurements of the $b$-value for lines of each 
species, the kinetic temperature, $T_k$, and the turbulent velocity, $v_{turb}$,
can be extracted,

\begin{equation}
b^2 = \frac{2kT_k}{m} + 2 v_{turb}^2.
\end{equation}

\noindent The 
parameters $k$ and $m$ are the Boltzmann constant and the mass of the atom
observed.  The measured $b$-values for
each species for individual velocity components along a given LOS are listed in 
Table~\ref{fit_results}.  The $b$-values for Li are generally larger than those 
for K, with the exception of the LOS toward X~Per and $\zeta$~Oph.  If 
$v_{turb}$ is assumed to be negligible, an upper 
limit can be placed on the kinetic temperature.  The upper limit on $T_k$ 
for all sight lines studied here ranges from 100 K to 2,700 K, values 
not unexpected for diffuse interstellar clouds.  The kinetic temperature and 
turbulent velocity can be determined simultaneously, since there 
are two independent measures of the $b$-value.  The calculated values of 
$T_k$ and $v_{turb}$ for each velocity component are given in Table~\ref{Temp}.
The uncertainties are based on an assumed 30\% error in the $b$-values for both
species.

In order to arrive at a solution for $v_{turb}$ for the second component 
toward $o$~Per, both components toward X~Per, and the single component toward 
$\chi$~Oph, extreme values of the $b$-value ($\pm$ 1-$\sigma$) were utilized. 
No error bars are quoted for $T_k$ or $v_{turb}$ for these components.  The 
derived temperatures for 
the two components toward $o$~Per are 98~$\pm$ 42~K and 900 K and those toward 
X~Per are 520 K and 788 K.  For the single component toward $\chi$~Oph, a 
temperature of 220 K is inferred.  The two components toward $\zeta$~Oph yield 
temperatures of 61 $\pm$ 27 K and 63 $\pm$ 27 K and toward 20~Aql the 
temperatures are 287 $\pm$ 124 K and 770 $\pm$ 330 K.  The current 
measurements agree with previous determinations of $T_k$, to within the 
assumed uncertainties.  From H$_2$ excitation, Savage et al. (1977) deduced 
temperatures of 48 K toward $o$~Per, 46 K toward $\chi$~Oph, and 54 K 
toward $\zeta$~Oph. From C$_2$ excitation and a simple 
chemical model for CH, C$_2$, and CN (Federman et al. 1994), temperatures 
of 40 K toward $o$~Per, 20 K toward X~Per, 60 K toward $\chi$ 
and $\zeta$~Oph (Federman et al. 1994), and 50 $-$ 60 K toward 20~Aql 
(Federman, Strom, \& Good 1991; Hanson, Snow, \& Black 1992; Knauth et al. 
2001) are derived.  From an analysis of C$_2$ excitation, Wannier et al. 
(1999) derived a temperature of 40 $\pm$ 20 for the LOS toward $o$~Per and 
Lambert, Sheffer, \& Federman (1995) derived a temperature of 20 $-$ 80 K for 
the interstellar clouds toward $\zeta$~Oph.  Howarth et al. (2002) found 
$T_k$ of 500 $\pm$ 130 K and 830 $\pm$ 125 K toward $\zeta$~Oph.  Their 
higher temperatures are primarily due to their larger derived $b$-values for 
Li~{\small I}.

The turbulent velocities were determined for each velocity component as well
as $T_k$.  Our calculated 
values of $v_{turb}$ for both interstellar clouds toward $\zeta$~Oph are 0.38 
$\pm$ 0.20 and 0.30 $\pm$ 0.16 km s$^{-1}$.  These values agree well with 
previous determinations, 0.39$^{+0.04}_{-0.10}$ and 0.33$^{+0.37}_{-0.16}$ km 
s$^{-1}$, based on UHR spectra of CH and CN (Crawford et al. 1994).  This 
agreement with an independent measure places additional confidence in 
$v_{turb}$ obtained for the other sight lines.  With the exception of the 
second component toward $o$~Per and the LOS toward X~Per, the turbulent 
velocities are substantially less than 1 km s$^{-1}$.  For further comparison,
$v_{turb}$ was calculated assuming $T_k$ = 100 K, a typical value for diffuse 
clouds.  A comparison between 
$v_{turb}$ and the sound speed shows that the turbulence is subsonic for all 
sight lines.  The inclusion of molecular gas decreases the sound speed 
slightly and in some instances may lead to sonic turbulence. 

\section{Interstellar $^7$Li/$^6$Li Ratios - the rule}

With the exception of one cloud along the line of sight to $o$~Per
and possibly the line of sight toward $\zeta$\,Oph, our
lower limits and the few other published $^7$Li/$^6$Li ratios are
consistent with an expectation that interstellar Li should be
similar to the Solar System ratio of $^7$Li/$^6$Li = 12.3 (Anders \& Grevesse 
1989). This expectation is based on observational evidence that young stars and
interstellar gas have a composition quite similar to that of the Sun for
elements whose synthesis is traceable to the same sites likely to
control the $^7$Li and $^6$Li input into the interstellar
medium.  Beryllium and $^6$Li are both products of spallation between
Galactic cosmic rays and interstellar nuclei. The stellar abundances
of Be in local stars younger than the Sun is very similar to the
Solar System abundance (Boesgaard et al. 2001).  Oxygen, a product of
SN II which may also make $^7$Li by the $\nu$-process, has
very similar abundances in the Sun, young stars, and the interstellar
medium (Allende Prieto, Lambert, Asplund 2001). Carbon, partly a product of AGB
stars which may also synthesize $^7$Li, also shares a similar
abundance with Sun, young stars, and the local interstellar
medium (Allende Prieto, Lambert, \& Asplund 2002).  Another line of evidence
to support the expectation of similar isotopic ratios among the
clouds and the Solar System is that stars of solar
metallicity show very similar abundance ratios for elements created by 
different nucleosynthetic processes despite differences in
age and birthplace (Edvardsson et al. 1993; Reddy et al. 2002).

The expectation applies to the total Li abundance too. Since
about 90\% of interstellar Li is depleted onto or into
grains, the K/Li ratio is considered. Identification of this ratio
with the elemental ratio assumes depletion of Li and K,
both alkali atoms, is similar. Table~\ref{K/Li} shows that
seven of the nine clouds have a K/Li ratio close to the
Solar System ratio of 61.6: the mean of the 7 clouds is
62.4 and individual measurements are each consistent with the
Solar System value to within their mutual uncertainties.  Our previous
observations toward $\zeta$~Per (Knauth et al. 2000) yield a K/Li ratio of 52.6
$\pm$ 3.4.  The average of eight additional K/Li ratios from Welty \& Hobbs 
(2001) is 59, assuming that a standard extinction curve applies for all sight 
lines.  With the exception of 20~Aql, all K/Li ratios are consistent with the 
Solar System value.

The interstellar clouds, toward stars studied here, have depletion indices ranging
from 0.7 to 1.5 dex (assuming solar Li and K abundances).
The sight line with the largest depletion, X~Per, is the most reddened 
direction.  The other two clouds, the pair toward 20 Aql, have K/Li  
$\simeq$ 116, a value about twice the Solar System ratio.  Several speculations
may be offered for the anomalous ratios toward 20~Aql: (i) the spectrum of the
ionizing radiation is different from that assumed; (ii) Li (relative
to K) is more depleted onto grains; or (iii) the gas
may have been mixed with  ejecta --now cold -- from the supernova
thought to be responsible for Radio Loop I (Hayakawa et al. 1979; 
Sembach, Savage, \& Tripp 1997). Factors of two
spreads in abundance ratios  are not uncommon;
Welty \& Hobbs (2001) discuss a collection of `discrepant' clouds toward the
Sco-Oph region.

\section{Interstellar $^7$Li/$^6$Li Ratios - the exception?}

One of the clouds toward $o$~Per has an exceptional $^7$Li/$^6$LI
ratio of 2.1 $\pm$ 1.1.
This is clearly shown by spectra taken with two different spectrographs.
Our previous lower resolution data shown in Figure~\ref{nature} clearly
shows that a Solar System ratio cannot apply to either interstellar
cloud toward $o$~Per. From a visual examination of the data, both components 
are in almost a one-to-one correspondence and not in the two-to-one 
relation expected from atomic physics.  
The apparent disagreement in the $^7$Li/$^6$Li isotope ratio for the second 
component, between our current and previous measurements, could be due 
to an undetected CCD defect and/or the sole use of the observational errors 
in determining the uncertainties of the profile synthesis.

In interpreting the low isotope ratio, possible scenarios 
include: (i) an interloper providing $^7$Li at the 
wavelength of the $^6$Li line; (ii) fractionation of the Li
isotopes leading to an incorrect isotopic ratio for Li atoms in the
gas; or (iii) a local change in the isotopic ratio arising from synthesis
of Li isotopes.  

Thinner clouds than detectable from our 4044 \AA\ spectra
are revealed by spectra of the 7699 \AA\ K~{\small I}
and Na D lines. If the absorption attributed to $^6$Li
is in fact $^7$Li with a column density of
about 2 $\times$ 10$^8$ cm$^{-2}$, a typical diffuse cloud
should have a K~{\small I} column density of about 4 $\times$ 10$^{10}$
cm$^2$, and a detectable line at  7699 $\mbox{\AA}$. Its 4044 \AA\
counterpart would have an $W_{\lambda}$ $<$ 0.1 m\AA\ and be indistinguishable
from the noise in the spectrum shown in Figure~\ref{ofit1} (top panel). The 
fifth cloud shown by Welty \& Hobbs has a column density of 3.9 $\times 
10^{10}$ cm$^{-2}$, but it is displaced to the red from the parent $^7$Li
line by about 4.7 km s$^{-1}$ not the by 7.1 km s$^{-1}$ that is
the isotopic shift.  However, inclusion of this red component into the $^7$Li
profile synthesis does not resolve the low isotope ratio.  

Even thinner clouds are detectable in the Na D lines. 
Welty \& Hobbs (2001) found two additional clouds to the red of
the two in Figure 1. The velocities are +10.6 and +14.0 km s$^{-1}$
with Na~{\small I} column densities of
4.9 $\times$ 10$^{10}$  cm$^{-2}$ and 2.5 $\times$ 10$^{10}$
cm $^{-2}$, respectively (D. Welty 2001, private communication).  The 10.6 km 
s$^{-1}$ cloud's stronger $^7$Li line needs to be at +11.1 km s$^{-1}$ to be 
mistaken for $^6$Li in the weaker of our two clouds.  However, if the +10.6 km 
s$^{-1}$ cloud has a normal Na~{\small I}/Li~{\small I} ratio, the expected 
Li~{\small I} column density is almost 2 orders of magnitude less than
that required to account for the $^6$Li.  Incorporation of these
two additional velocity components into the profile synthesis yields
a reasonable fit to the data and obtains $^7$Li/$^6$Li ratios that are all
consistent with the Solar System value.  Unfortunately, this scenario
solves the $^6$Li-problem by replacing it with another one -
how to account for the very discrepant Na~{\small I}/Li~{\small I}
and K~{\small I}/Li~{\small I} ratios yet normal Na~{\small I}/K~{\small I}
ratios. 

Fractionation of the Li isotopes is an improbable
scenario on several accounts. Even if no $^6$Li is removed
from the gas, the total Li abundance must drop by a factor of
about 4, which would place the cloud off the
tight relation in Figure~\ref{livsk}, unless K is also removed.
In addition, we have been unable to identify a plausible
fractionation process:   the predicted abundances of
Li-containing molecules (Kirby \& Dalgarno 1978) are orders of
magnitude too low; ionization and recombination are
isotopically insensitive; gas-dust grain processes
would seem not to be dependent on the mass difference
between the isotopes.\footnote{If the grains originate in
circumstellar atmospheres of red giants, they will be
Li-poor in the main. Then, evaporation of grains will
produce a local dilution of the Li/K ratio but no change in
the $^7$Li/$^6$Li ratio.}  Therefore, it is unlikely that $^7$Li is hiding
in another neutral species.

A close correspondence between the exceptional $^7$Li/$^6$Li
ratio and that resulting from either spallation of C, N, and O 
or $\alpha$ + $\alpha$ fusion reactions encourages speculation
that the ratio reflects freshly synthesized lithium. The
speculation faces at least two hurdles. (i) Why is the K/Li ratio
of the exceptional cloud that of a normal cloud where local
synthesis of Li is not invoked? (ii) What is the energy source
for the particles causing the spallation and fusion reactions?

Omicron Per is a member of the Per OB2 Association
and the line of sight lies close to IC 348, an active region of
star formation. The association, as would be expected, has
experienced supernovae. X Per is a O9.5Ve star orbiting a
neutron star (Delgado-Mart\'{\i} et al. 2001) and $\xi$ Per
is a runaway star (Hoogerwerf, de Bruijne, \& de Zeeuw 2001). Consideration 
of the chemistry of OH and HD led Federman et al. (1996) to the
conclusion that the interstellar clouds toward $o$ Per 
are permeated by an order of magnitude higher
cosmic ray flux than clouds seen toward other stars in Per OB2. 
The enhanced cosmic rays could indicate that we are looking at a young
superbubble where dilution has not yet occurred.

The energetics required to synthesize the observed amount of Li toward $o$~Per
may reveal vital information for understanding Li production in interstellar
space.  Assuming spherical geometry and using $N_{tot}$(H) and $n$ from 
Table~\ref{proton}, we find that 2.2 solar masses of hydrogen are present in
the interstellar clouds toward $o$~Per.  Assuming that the clouds have the 
Solar System Li abundance (Li/H$_{\odot}$ = 20.5 $\times$ 10$^{-10}$) implies 
that there are 5.4 $\times$ 10$^{48}$ atoms of Li in the clouds.  Approximately 
3 ergs of energy are needed to create a single atom of Li (Ramaty et al. 
1996).  Therefore, to reproduce the entire Li content in the clouds toward
$o$~Per requires approximately 1\% of the energy output of a typical supernova
explosion.  Models of superbubbles (Parizot \& Drury 1999) show that the Li 
production per SN~{\small II} is approximately 10$^{50}$ atoms.  The 
superbubble picture may apply for the interstellar clouds toward $o$~Per.
We note that Li is not produced solely through supernova explosions.  

Studies show large variable x-ray fluxes in IC~348 (Preibisch, Zinnecker, \& 
Herbig 1996) that are attributed to flares produced in 
T Tauri stars (Preibisch, Neuh$\ddot{\rm a}$user, \& Alcal\'{a} 1995).  The 
total x-ray luminosity is on the order of 10$^{32}$ 
erg s$^{-1}$ (Preibisch et al. 1996).  If all 116 x-ray sources (Preibisch et 
al. 1996) contribute, with flares lasting on the order of 10 hours, the total
energy produced is 4 $\times$ 10$^{38}$ ergs, enough for about 10$^{38}$ 
atoms of $^6$Li.  This calculation assumes that all flare energy 
goes into the creation of $^6$Li.  
The amount of $^6$Li that can be created by stellar flares is about 10 
orders of magnitude less than that needed to reproduce the entire $^6$Li 
content of the interstellar clouds toward $o$~Per.  X-ray flares with energies 
on the order of 10$^{38}$ ergs have been reported (Preibisch et al. 1995).  
Even in these extreme cases, $^6$Li production is negligible; therefore x-ray 
flares cannot produce the observed $^6$Li abundance.  A further complication 
is that this process is not isotope selective since $^7$Li is formed as well.

\section{Constraints on Stellar Source of $^7$Li}
The isotope ratios given in Table~\ref{fit_results} provide important constraints on
stellar production pathways for $^7$Li (Reeves 1993; Steigman 1993).  
The abundance of $^7$Li arises from several processes: BBN, GCR spallation 
reactions, and stellar production mechanisms.  During the early Universe, BBN
supplied about 10\% of the present $^7$Li and a negligible amount of $^6$Li. 
GCR spallation reactions contribute another 10~$-$~25\% of $^7$Li and account
for the present abundance of $^6$Li.  Although the nature of the stellar source
of $^7$Li is still unclear, it plays the most important role in Li production
today. 

There has been much debate over which stellar source dominates $^7$Li
production throughout the lifetime of the Galaxy.  As noted in the Introduction,
the various stellar sources of $^7$Li are thought to be primarily 
AGB and RGB stars, SN II, and possibly novae.  Although AGB have been observed
with the largest Li abundances, recent models of 
Galactic chemical evolution have all but eliminated AGB stars as contributors 
to the Galactic $^7$Li abundance (Charbonnel \& Balachandran 2000; 
Romano et al. 2001).  As for the $\nu$-process in SN~{\small II}, the yields of
$^{11}$B and therefore $^7$Li have been called into question.  Alibes,
Labay, and Canal (2001) suggest a reduction by a factor of 2 in the yields of
Woosley and Weaver (1995).  A problem with the source being novae is the lack of 
observational evidence (Andrea, Drechsel, \& Starrfield 1994; Matteucci, 
D'Antona, \& Timmes 1995).  Since stars have the 
greatest effect on the $^7$Li production today, constraints are needed to 
isolate the stellar source(s).   

Knowledge of the $^6$Li abundance yields the amount of $^7$Li
produced via GCR spallation because the isotope ratio predicted from
these reactions (R$_{76,GCR}$) is known with sufficient accuracy (Ramaty et al.
1997; Lemoine et al. 1998).  The isotope ratio used here is R$_{76,GCR}$ = 1.6,
which is midway between the values presented by Ramaty et al. (1997).   If the 
$^7$Li abundance in halo stars represents the primordial abundance (Ryan et al. 
2000, 2001; Suzuki et al. 2000), then the BBN contribution is fairly well
known.  Additional support for Li in halo stars representing the
primordial abundance comes from recent determinations of the $^7$Li/$^6$Li 
ratio (Smith, Lambert, \& Nissen 1993; Hobbs \& Thorburn 1994, 
1997; Hobbs, Thorburn, \& Rebull 1999).  These measurements show that there 
has been little or no nuclear depletion of $^7$Li during the lifetimes 
($\sim$ 12 Gyr) of these stars.  Observations of $^6$Li in halo 
stars also yield a measure of the contamination of the primordial $^7$Li
through GCR spallation reactions.   Recent theoretical 
determinations of the primordial Li abundance find $A$(Li)$_p$ = 
2.09 (Suzuki et al. 2000), or (Li/H)$_p$ = 1.23 
$\times$ 10$^{-10}$.  The following equation allows the removal of the 
``known" sources of $^7$Li from the present interstellar abundance.  
$^7$Li$_{stars}$ represents the combined contribution from all stellar 
sources (AGB, RGB, SN~{\small II}, \& novae),

\begin{equation}
^7{\rm Li}_{stars}~=~^7{\rm Li}_{ISM}~-~^6{\rm Li}_{ISM}~*~{\rm
R}_{76,GCR}~-~^7{\rm Li}_{p}.
\end{equation}

\noindent  Also included in $^7$Li$_{stars}$ is
the amount of destruction
by stellar astration.

The Li abundances corrected for our estimates of depletion onto grains and the 
amount of $^7$Li
produced through GCR spallation reactions are presented in 
Table~\ref{*source}.  The amount of $^7$Li produced by all stellar sources 
(using our depletion index) is similar, approximately $^7$Li/H $\sim$ 1 $\times$
10$^{-9}$ for all lines of sight.  A larger spread arises (-5.7 $\times$
10$^{-12}$ $-$ 1.5 $\times$ 10$^{-9}$) when utilizing a constant depletion 
index (Welty \& Hobbs 2001), suggesting that use of a constant depletion 
index may not be appropriate.  Since $^7$Li production through the various
stellar sources in our sample of sight lines is essentially constant, we infer
that one or two stellar processes dominate Li production in the solar
neighborhood.  This may not be unreasonable given the relatively 
small volume of the Galaxy studied. 

Comparison with other elements and isotopic ratios are needed as well.  A
recent study (Sofia, Meyer, \& Cardelli 1999) showed that several lines of 
sight toward Orion exhibit large interstellar Sn (an s-process element) 
abundances.   Sofia et al. (1999) attribute this Sn overabundance to AGB 
stars.   Since AGB stars and SN {\small II} explosions also generate Li, 
simultaneous observations of Li, s-process, and r-process elements are needed 
to constrain the amounts of Li production by these two stellar sources.   
Comparisons of r-process and s-process elements in stars are 
important for determining which stellar processes contribute to the 
production of these elements.  Knowledge of the $^{12}$C/$^{13}$C and oxygen 
isotope ratios is useful in constraining the production of Li in RGB stars.  
For instance, there could be a low $^{12}$C/$^{13}$C ratio in stars with 
high Li abundance (Sackmann \& Boothroyd 1999), but Charbonnel and Balachandran
(2000) predict that the high $^7$Li abundances will be followed by the rapid 
destruction of both $^7$Li and $^{12}$C.  
Therefore, stellar and interstellar studies are required to disentangle the
contributions from the different stellar sources.

\section{Conclusions}

The interstellar Li abundance and the $^7$Li/$^6$Li isotope ratio were measured
toward several bright stars in the solar neighborhood.  These observations form
the highest resolution (R $\sim$ 360,000) survey of interstellar Li to date.
From these data it appears that the Li abundance has not significantly changed
since the formation of the Solar System and that the Solar System value of 
$\sim$ 12 for the $^7$Li/$^6$Li ratio represents most of the gas in the solar 
neighborhood.  One exception is the LOS toward $o$~Per.  This sight 
line exhibits a low $^7$Li/$^6$Li ratio of about 2, the value expected from 
models of GCR spallation reactions, which is consistent with our
previous determination (Knauth et al. 2000) from lower
resolution observations.  

If a value of about 12 was the starting condition for the gas 
toward $o$~Per, the decrease in the ratio to its 
present value requires an increase in the Li abundance by an order of 
magnitude.  However, an enhancement in the total Li abundance is 
not seen.  The elemental K/Li ratio for this gas is not unusual either, 
although K and Li are produced via different nucleosynthetic pathways.  The 
most likely scenario for the ISM toward $o$~Per is that the initial Li was
enhanced through GCR spallation reactions occurring as a result of recent
supernovae grouped in a superbubble.  The cloud containing a low isotope
ratio contains about 20\% of the total Li content of the clouds.  Therefore,
determination of a total Li abundance may mask the true Li abundance for
each interstellar cloud. Although the supernova scenario explains 
many of the observations, this hypothesis does not address the similarity of 
the K/Li ratio with results for other sight lines.  
Fractionation cannot account for the low
$^7$Li/$^6$Li ratio because the abundance of LiH is too small.  There is also
evidence for enhanced x-ray activity in the cluster IC~348 caused by stellar
flares in T Tauri stars (Preibisch et al. 1996).  While $^6$Li can be produced
during stellar flares (e.g., Deliyannis \& Malaney 1995; Ramaty et al. 2000b),
the enhanced flaring activity in OB associations was found to play
a negligible role.  More typical $^7$Li/$^6$Li
ratios can be obtained through the inclusion of extra velocity components that 
are detected in Na~{\small I} absorption at 5895 \AA\ but not seen in 
K~{\small I} absorption at 7699 $\mbox{\AA}$, although this comes at the 
expense of highly disparate Na~{\small I}/Li~{\small I} and 
K~{\small I}/Li~{\small I} 
ratios.  Therefore, the exceptional isotopic ratio toward $o$~Per remains
a puzzle. Further observations of other stars in or near IC~348 are
needed to probe the extent of the region containing a low $^7$Li/$^6$Li ratio. 
 
Observations were made of X~Per, a nearby x-ray binary.  It is believed that
X~Per and other x-ray binary systems are formed after a supernova explosion 
(e.g., Delgado-Mart\'{i} et al. 2001).  The direction toward X~Per shows 
almost an order of magnitude smaller Li and K abundances than toward other 
sight lines.  These smaller abundances suggest enhanced depletion along the
LOS.    Unfortunately, the SNR of the X~Per spectrum was too low to allow
a useful probe of the region containing the low isotope ratio.

The stars in Sco OB2 were studied since they closely matched the situation
present toward $o$~Per.  These stars are in an OB association where star 
formation is occurring in the $\rho$ Oph molecular cloud.  There is 
evidence of a past supernova because $\zeta$ Oph is a runaway star 
(Hoogerwerf et al. 2001).  Why is there no evidence for 
enhanced $^6$Li production in the presence of active star formation 
($\rho$~Oph molecular cloud) and past SN~{\small II} explosions?  Is the LOS
toward $o$~Per really that rare?  More observations are necessary to determine 
if we are experiencing a selection effect; to date, only three sight lines in 
each OB association have been studied with sufficient precision. 

An additional result of our study involves the LOS toward 20~Aql, which 
resides far from active star-forming regions.  The $^7$Li/$^6$Li ratio is 
similar to the Solar System value, but the K/Li ratio is almost double the
expected value.  The K/H abundance is similar to the abundance
found for other sight lines, indicating that there is less Li.
The LOS toward 20~Aql lies on the edge of Radio Loop I (Hayakawa et al. 1979;
Sembach, Savage, Tripp 1997), a superbubble.  
However, the timescale needed to destroy Li through proton burning
is significantly longer than the age of bubble.  Another explanation
is required to account for the low Li/H abundance.  The low Li abundance
toward 20~Aql could be evidence for Li dilution in an old superbubble
(Parizot \& Drury 1999; Knauth et al. 
2000).  Observations of Li~{\small I} and K~{\small I} toward other stars 
in the vicinity of 20~Aql are necessary to gain further insight.

We place additional interstellar constraints on the $^7$Li produced 
by a stellar source.  Our results indicate that 
there is essentially a constant production of $^7$Li in stars in the solar 
neighborhood, suggesting one or two processes dominate.  Observations of 
interstellar and stellar $^7$Li, s-process, and r-process elements can
be of great value in constraining the Li production by the various stellar 
sources.  

For future studies of interstellar Li, knowledge of the K~{\small I} 
abundance in conjunction with the molecular content can be used to find sight 
lines that contain observable amounts of Li.  For instance, the results of 
this survey suggest other targets in Sco OB2.  These stars, which
are fainter than X~Per, can be observed at lower resolving power (R $\sim$ 
180,000) because their interstellar spectra reveal simple velocity structure. 
Preliminary results for these stars indicate that the $^7$Li/$^6$Li ratios are
similar to the Solar System value.  Thus, lines of sight with low 
$^7$Li/$^6$Li ratios appear to be rare.  Observations of stars in the vicinity
of IC~348 (in Per OB2) are planned to help clarify the picture for gas toward
$o$~Per.  In addition, data from the $Hubble~Space~Telescope$ on 
the interstellar $^{11}$B/$^{10}$B ratio toward stars in Per OB2 are being
analyzed.  The interstellar abundance of $^9$Be and the 
$^{11}$B/$^{10}$B ratio (Hebrard et al. 1997; Lambert et al. 1998) are crucial
to our understanding of Li evolution.
Knowledge of $^9$Be and $^{10}$B further constrain the amount 
of GCR spallation occurring along the line of sight, since GCR spallation is 
the only production route for these light elements.  Observations of 
interstellar $^9$Be are not yet feasible because its gas phase abundance is
extremely low ($^9$Be/H $\leq$ 7 $\times$ 10$^{-13}$; Hebrard et al. 1997).  
The amount of $^{11}$B also yields constraints on SN~{\small II} explosions 
because about 50\% of 
$^{11}$B is thought to arise from neutrino-induced spallation reactions 
(Woosley et al. 1990; Woosley \& Weaver 1995).    Only through a 
more complete study of the abundances for a variety of atomic species will 
the light element puzzle be resolved.  

\acknowledgments

\acknowledgments  The McDonald Observatory data presented here 
formed the basis for David Knauth's Ph.D. dissertation.  It is a pleasure to 
thank the excellent staff at McDonald Observatory, especially David Doss, 
for assistance with the instrumental setup.  We also thank J\'{a}nos 
Zsarg\'{o} for the use of his code.  This research made use of the Simbad 
database, operated at CDS, Strasbourg, France. This research was supported 
in part by NASA LTSA grant NAG5-4957.

\clearpage

\newpage
\clearpage

\newpage
\clearpage

\newpage
\clearpage

\newpage
\clearpage

\newpage
\clearpage

\newpage
\clearpage

\newpage
\clearpage

\newpage
\setcounter{table}{0}
\begin{deluxetable}{ccccccccc}
\tablecolumns{9}
\tablewidth{0pt}
\tablecaption{Stellar Data \label{stars-table}}
\tablehead{ 
\colhead{Star} & \colhead{Name} & \colhead{Spectral Type$^{~{\rm a}}$} & 
\colhead{$V^{{\rm a}}$} & 
\colhead{$B-V^{{\rm a}}$} & \colhead{$l^{{\rm a}}$} & 
\colhead{$b^{{\rm a}}$} & \colhead{$d^{{\rm b}}$} & \colhead{$E(B-V)^{{\rm c}}$} \\
\colhead{} & \colhead{} & \colhead{} & \colhead{[mag]} & \colhead{[mag]} & \colhead{[deg]} & \colhead{[deg]} & \colhead{[pc]} &
\colhead{[mag]}}
\startdata
HD 5394 & $\gamma$~Cas & B0 IVe & 2.39 & -0.10 & 123.58 & -2.15 & 188 & 0.14 \\
\\
HD 23180 & $o$~Per  & B1 III & 3.86 & 0.02 & 160.36 & -17.74 & 453 &  0.26 \\
\\
HD 22951 & 40~Per & B0.5 V  & 4.98 & -0.05 & 158.92 & -16.70 & 283 & 0.23  \\
 \\
HD 24534 & X~Per & O9.5 V & 6.10 & 0.29 & 163.08 & -17.14 & 826 & 0.56 \\
\\
HD36861 & $\lambda$~Ori & O8 IIIf & 3.39 & -0.19 & 195.05 & -12.00 & 324 & 0.12 \\
\\
HD 87901 & $\alpha$~Leo & B7 V & 1.35 & -0.11 & 226.43 & 48.93 & 24 & $\ldots$ \\
\\
HD 116658 & $\alpha$~Vir & B1 III-IV$^+$ & 1.04 & -0.13 & 316.11 & 50.84 & 80 & $\ldots$ \\
\\
HD 148184 & $\chi$~Oph  & B2 Vne & 4.42 & 0.28 & 357.93 & 20.68 & 150 & 0.49 \\
 \\
HD 149757 & $\zeta$~Oph & O9.5 V & 2.58 & 0.02 & 6.28  & 23.59 & 140 & 0.32 \\
\\
HD 179406 & 20~Aql & B3 V & 5.36 & 0.09 & 28.23  & -8.31 & 373 & 0.33 \\
\enddata
\footnotesize{
\tablenotetext{a}{Simbad database, operated at CDS, Strasbourg, France.}
\tablenotetext{b}{Perryman et al. (1997).}
\tablenotetext{c}{Papaj, Krelowski, \& Wegner (1991).}}
\end{deluxetable}
\clearpage

\newpage
\begin{deluxetable}{cccccc}
\tablecolumns{6}
\tablewidth{0pt}
\tablecaption{Li~{\small I} and K~{\small I} Parameters\label{IPfw-table}}
\tablehead{ 
\colhead{Species} & \colhead{IP$^{{\rm a}}$} & 
\colhead{$\lambda_{fs}^{{\rm b},{\rm c}}$} & \colhead{$f_{fs}^{{\rm a}}$} &
\colhead{$\lambda_{hfs}^{{\rm b},{\rm c}}$} & \colhead{$f_{hfs}^{{\rm d}}$} \\
\colhead{} & \colhead{[eV]} & \colhead{[$\mbox{\AA}$]} & \colhead{} & \colhead{[$\mbox{\AA}$]} &  
\colhead{}}
\startdata
$^7$Li~{\small I} & 5.392 & 6707.764 & 0.4946 & 6707.757 & 0.18550 \\
 & & & & 6707.769 & 0.30910 \\
\\
 & & 6707.915 & 0.2473 & 6707.908 & 0.09270 \\ 
 & & & & 6707.920 & 0.15460 \\
\\
$^6$Li~{\small I} & 5.392 & 6707.924 & 0.4946 & 6707.922 & 0.16490 \\
 & & & & 6707.925 & 0.32970 \\
\\
 & & 6708.075 & 0.2474 & 6708.073 & 0.08243 \\
 & & & & 6708.076 & 0.16490 \\
\\
K~{\small I} & 4.341 & 4044.143$^{~{\rm a}}$ & 0.006089 & $^{{\rm e}}$ &
$^{{\rm e}}$ \\
\enddata
\footnotesize{
\tablenotetext{a}{Morton (1991).} 
\tablenotetext{b}{Sansonetti et al. (1995).} 
\tablenotetext{c}{$fs$ $-$ Fine structure, $hfs$ $-$ Hyperfine structure.} 
\tablenotetext{d}{Welty, Kulkarni, \& Hobbs (1994).}
\tablenotetext{e}{Hyperfine splitting of K~{\small I} $\lambda$4044 was found 
to be negligible.}} 
\end{deluxetable}
\clearpage

\newpage
\begin{deluxetable}{cccccc}
\tablecolumns{6}
\tablewidth{0pt}
\tablecaption{UHR Li~{\small I} and K~{\small I} Observations \label{obs-table}}
\tablehead{ 
\colhead{Star} & \colhead{Species} & \colhead{Observatory$^{\rm a}$} &
\colhead{Date} & \colhead{Exposure Time} & \colhead{SNR/pixel$^{\rm b}$}  \\
\colhead{} & \colhead{} & \colhead{} & \colhead{} & \colhead{[s]} & \colhead{} }
\startdata
$\gamma$~Cas & Li~{\small I} & McD & 08/02/1998 $-$ 08/08/1998 & 19,200 & 1000 \\
40~Per & Li~{\small I} & McD & 08/11/1999 $-$ 08/16/1999 & 14,400 &  270 \\
 &  & McD & 09/15/2000 $-$ 09/21/2000 & 43,200 & 470 \\
$o$~Per & Li~{\small I} & McD & 12/30/1998 $-$ 01/05/1999 & 103,926 & 720 \\
 &  & McD & 12/23/2000 $-$ 12/30/2000 & 81,827 & 600 \\
 & K~{\small I} & McD & 07/24/1999 $-$ 07/28/1999 & 7,200 & 100 \\
 &  & McD & 08/11/1999 $-$ 08/16/1999 & 27,000 & 290 \\
 &  & McD & 12/23/2000 $-$ 12/30/2000 & 28,800 & 200 \\
X~Per & Li~{\small I} & McD & 12/28/1999 $-$ 12/30/1999 & 45,000 & 150 \\
 &  & McD & 09/15/2000 $-$ 09/21/2000 & 75,600 & 290 \\
$\lambda$~Ori & Li~{\small I} & McD & 12/30/1998 $-$ 01/05/1999 & 52,200 & 420 \\    
 &  & McD & 12/28/1999 $-$ 12/30/1999 & 21,600 & 370 \\
$\alpha$~Leo & Li~{\small I} & McD & 12/30/1998 $-$ 01/05/1999 & 37,800 & 1050 \\
 &  & McD & 12/28/1999 $-$ 12/30/1999 & 10,800 & 850 \\
 & K~{\small I} & McD & 12/23/2000 $-$ 12/30/2000 & 4,500 & 170 \\
$\alpha$~Vir & Li~{\small I} & McD & 12/23/2000 $-$ 12/30/2000 & 12,127 & 470 \\
 & K~{\small I} & McD & 05/11/2000 $-$ 05/13/2000 & 13,200 & 320 \\
o~Sco & K~{\small I} & AAO & 06/19/1995 $-$ 06/20/1995 & 2700 & 130 \\
$\rho$~Oph & K~{\small I} & AAO & 06/19/1995 $-$ 06/20/1995 & 5700 & 100 \\
$\chi$~Oph & Li~{\small I} & AAO & 06/19/1995 $-$ 06/20/1995 & 27900 & 320 \\
 &  & McD & 09/15/2000 $-$ 09/21/2000 & 9,000 & 270 \\
 & K~{\small I} & AAO & 06/19/1995 $-$ 06/20/1995 & 5700 & 160 \\
 &  & McD & 05/11/2000 $-$ 05/13/2000 & 50,400 & 150 \\
$\zeta$~Oph & Li~{\small I} & AAO & 06/10/1994 $-$ 06/12/1994 & 64800 & 415 \\
  & & McD & 08/02/1998 $-$ 08/08/1998 & 13,500 & 500 \\
  & & McD & 09/15/2000 $-$ 09/21/2000 & 12,600 & 430 \\
 & K~{\small I} & AAO & 06/19/1995 $-$ 06/20/1995 & 4800 & 200 \\
 & & McD & 07/24/1999 $-$ 07/28/1999 & 44,700 & 580 \\
HD~154090 & K~{\small I} & AAO & 06/19/1995 $-$ 06/20/1995 & 3600 & 140 \\
HD~165024 & K~{\small I} & AAO & 06/19/1995 $-$ 06/20/1995 & 2700 & 160 \\
$\mu$~Sgr & K~{\small I} & AAO & 06/19/1995 $-$ 06/20/1995 & 5800 & 130 \\
20~Aql & Li~{\small I} & McD & 08/02/1998 $-$ 08/08/1998 & 51,728 & 410 \\
 &  & McD & 08/11/1999 $-$ 08/16/1999 & 43,200 & 410 \\
 &  & McD & 09/15/2000 $-$ 09/21/2000 & 18,000 & 250 \\
 & K~{\small I} & McD & 07/24/1999 $-$ 07/28/1999 & 32,400 & 100 \\
 &  & McD & 08/11/1999 $-$ 08/16/1999 & 55,453 & 200 \\
\enddata
\footnotesize{
\tablenotetext{a}{McD $-$ McDonald Observatory; AAO $-$ Anglo-Australian 
Observatory.}
\tablenotetext{b}{Final SNR per resolution element from combined binned
spectrum: $o$~Per $-$ Li~{\small I} 3040, K~{\small I} 1050; 40~Per $-$
Li~{\small I} 1500; X~Per $-$ Li~{\small I} 930; $\lambda$~Ori $-$ Li~{\small
I} 1440; $\chi$~Oph $-$ Li~{\small I} 1130, K~{\small I} 330; $\zeta$~Oph $-$
Li~{\small I} 2600, K {\small I} 1500; 20~Aql $-$ Li~{\small I} 1725,
K~{\small I} 780; $\gamma$ Cas $-$ Li~{\small I} 2825; $\alpha$~Leo $-$
Li~{\small I} 3675, K~{\small I} 480; $\alpha$ Vir $-$ Li~{\small I} 1330,
K~{\small I} 900.}}
\end{deluxetable}
\clearpage

\newpage
\begin{deluxetable}{cccc}
\tablecolumns{4}
\tablewidth{0pt}
\tablecaption{Li~{\small I} and K~{\small I} Upper Limits\label{UL-table}}
\tablehead{ 
\colhead{Star} & \colhead{Species} & \colhead{$W_{\lambda}$} & 
\colhead{$N$}  \\
\colhead{} & \colhead{} & \colhead{[m$\mbox{\AA}$]} & \colhead{[cm$^{-2}$]}}
\startdata
 $\alpha$~Vir & $^7$Li~{\small I} & $\leq$ 0.11 & $\leq$ 5.60 $\times$ 10$^8$ \\
 & K~{\small I}  & $\leq$ 0.18 & $\leq$ 2.04 $\times$ 10$^{11}$\\

$\alpha$~Leo & $^7$Li~{\small I}  & $\leq$ 0.05 & $\leq$ 2.55 $\times$ 10$^8$ \\
 & K~{\small I}  & $\leq$ 0.34 & $\leq$ 3.88 $\times$ 10$^{11}$ \\

40~Per & $^7$Li~{\small I}  & $\leq$ 0.11 & $\leq$ 5.60 $\times$ 10$^8$\\

$\lambda$~Ori & $^7$Li~{\small I} & $\leq$ 0.11 & $\leq$ 5.60 $\times$ 10$^8$\\

$\gamma$~Cas & $^7$Li~{\small I} & $\leq$ 0.07 & $\leq$ 3.55 $\times$ 10$^8$ \\
\enddata
\end{deluxetable}
\clearpage

\newpage
\begin{deluxetable}{ccccccccc}
\tablecolumns{9}
\tablewidth{0pt}
\tablecaption{Results of Profile Syntheses \label{fit_results}}
\tablehead{ 
\colhead{} &\colhead{Species} & \colhead{$V_{LSR}$$^{{\rm a}}$} & \colhead{$b$-value} & \colhead{$W_{\lambda}^{{\rm b}}$} &
\colhead{$W_{\lambda}$(FIT)} & \colhead{$N^{{\rm c}}$} & \colhead{$^7$Li/$^6$Li} & \colhead{$\chi^2$/$\nu$}  \\
\colhead{} & \colhead{} & \colhead{[km s$^{-1}$]} & \colhead{[km s$^{-1}$]} &
\colhead{[m$\mbox{\AA}$]} & \colhead{[m$\mbox{\AA}$]} & \colhead{[cm$^{-2}$]} & \colhead{} & \colhead{} }
\startdata
$o$~Per & & & & & & & \\
\\
 & $^7$Li~{\small I} & 4.19 & 0.53 & 0.12 $\pm$ 0.02 & 0.09 $\pm$ 0.02 & 
 4.7 $\pm$ 1.0 &   & 1.28 \\
 & & 7.38 & 2.53 & 0.47 $\pm$ 0.02 & 0.60 $\pm$ 0.02 & 30.6 $\pm$ 1.1 &    & 1.28 \\
 & $^6$Li~{\small I} & 4.19 & 0.53 & & 0.04 $\pm$ 0.02 & 2.2 $\pm$ 1.0 &  2.1 $\pm$ 1.1  & 1.28  \\
 & & 7.38 & 2.53 & & 0.08 $\pm$ 0.02 & 3.8 $\pm$ 1.0 &  8.1 $\pm$ 2.1   & 1.28 \\
& K~{\small I} & 3.96 & 0.30 & 0.14 $\pm$ 0.03 & 0.13 $\pm$ 0.03 & 1.46 $\pm$ 0.35 &  & 1.80 \\
 & & 6.98 & 0.92 & 0.74 $\pm$ 0.03 & 0.78 $\pm$ 0.03 & 8.96 $\pm$ 0.35 &  & 1.80 \\
\\
 X~Per & & & & & & & & \\
\\
 & $^7$Li~{\small I} & 5.45 & 1.20  & 0.63 $\pm$ 0.06 & 0.63 $\pm$ 0.06 & 32.0 $\pm$ 3.1 &    & 1.12 \\
 & & 7.85 & 1.15 & 0.35 $\pm$ 0.06 & 0.28 $\pm$ 0.06 & 14.2 $\pm$ 3.1 &    & 1.12 \\
 & $^6$Li~{\small I} & 5.45 & 1.20 & & $\le$ 0.10 & $\le$ 5.1 &  $\geq$ 6.3  & 1.12 \\
 & & 7.85 & 1.15 & & $\le$ 0.13 &  $\le$ 6.8 &  $\geq$ 2.1  & 1.12 \\
 \\
$\chi$~Oph & & & & & & & & \\
\\
 & $^7$Li~{\small I} & 0.887 & 1.27 & 0.93 $\pm$ 0.07 & 0.86 $\pm$ 0.07 & 44.0 $\pm$ 4.1 &    & 1.30
\\
 & $^6$Li~{\small I} & 0.887 & 1.27 & & 0.11 $\pm$ 0.07 & 5.8 $\pm$ 3.6 &  7.6 $\pm$ 4.8  & 1.30 \\
 & K~{\small I} & -0.30 & 0.46 & 1.03 $\pm$ 0.10 & 1.08 $\pm$ 0.10 & 12.7 $\pm$ 1.2 &    & 1.79 \\
\\
$\zeta$~Oph & & & & & & & & \\
\\
 & $^7$Li~{\small I} & -0.620 & 0.65 & 0.41 $\pm$ 0.02 & 0.41 $\pm$ 0.02 & 21.0 $\pm$ 1.0 &    & 1.81 \\
 & & 0.420 & 0.57 & 0.14 $\pm$ 0.02 & 0.22 $\pm$ 0.02 & 11.4 $\pm$ 1.0 &     & 1.81 \\
 & $^6$Li~{\small I} & -0.620 & 0.65 & & $\leq$ 0.03 & $\leq$ 1.8 & $\geq$ 11.7    & 1.81 \\
 & & 0.420 & 0.57 & & $\leq$ 0.02 & $\leq$ 1.0  &  $\geq$ 12.0   & 1.81 \\
 & K~{\small I} & -1.25  & 0.55 & 0.42 $\pm$ 0.03 & 0.47 $\pm$ 0.03 & 5.36 $\pm$ 0.35 &     & 3.62 \\
 & & -0.21 & 0.45 & 0.31 $\pm$ 0.03 & 0.25 $\pm$ 0.03 & 2.89 $\pm$ 0.34 &    & 3.62 \\
 \\
20~Aql & & & & & & & & \\
\\
 & $^7$Li~{\small I} & 2.06 & 1.06 & 0.67 $\pm$ 0.04 & 0.67 $\pm$ 0.04 & 34.0 $\pm$ 2.1 &    & 1.12 \\
 & & 3.35 & 1.46 & 0.24 $\pm$ 0.04 & 0.25 $\pm$ 0.04 & 12.8 $\pm$ 2.1 &    & 1.12 \\
 & $^6$Li~{\small I} & 2.06 & 1.06 & & $\leq$ 0.07 & $\leq$ 3.6 &  $\geq$ 9.4  & 1.12 \\
 & & 3.35 & 1.46 & & $\leq$ 0.04 & $\leq$ 2.3 & $\geq$ 5.6  & 1.12 \\
 & K~{\small I} & 1.60 & 0.75 & 1.30 $\pm$ 0.04 & 1.34 $\pm$ 0.04 & 15.6 $\pm$ 0.5 &    & 1.66 \\
 & & 3.21 & 0.79 & 0.59 $\pm$ 0.04 & 0.56 $\pm$ 0.04 & 6.40 $\pm$ 0.47 &    & 1.66 \\
\enddata
\footnotesize{
\tablenotetext{a}{Differences in velocities ($\sim$ 0.5 km s$^{-1}$) between 
Li~{\small I} and K~{\small I} lines, toward the same star, are attributed to
uncertainties in using the Li hollow cathode to determine wavelength 
calibration.}
\tablenotetext{b}{Measured $W_{\lambda}$ with 1-$\sigma$
observational uncertainty; No measured $W_{\lambda}$ for $^6$Li are reported.}
\tablenotetext{c}{ Note $-$ $N$(Li {\small I}) is in units of 10$^8$ cm$^{-2}$
and $N$(K {\small I}) is in units of 10$^{11}$ cm$^{-2}$.}} 
\end{deluxetable}
\clearpage

\newpage
\begin{deluxetable}{ccccc}
\tablecolumns{5}
\tablewidth{0pt}
\tablecaption{Hydrogen and Space Densities \label{proton}}
\tablehead{ 
\colhead{Star} & \colhead{$N$(H {\small I})$^{{\rm a}}$} & \colhead{$N$(H$_2$)$^{{\rm b}}$} 
& \colhead{$N_{tot}$(H)$^{{\rm c}}$} &
\colhead{$n^{{\rm d}}$}  \\
 \colhead{} & \colhead{[10$^{20}$ cm$^{-2}$]} & \colhead{[10$^{20}$ cm$^{-2}$]}
  & \colhead{[10$^{21}$ cm$^{-2}$]} & \colhead{[cm$^{-3}$]}}
\startdata
40~Per & 11.0 $\pm$ 4.7 & 2.88 $\pm$ 1.23  & 1.68 $\pm$ 1.01 & $\ldots$ \\
$o$~Per & 6.61 $\pm$ 1.38 & 4.07 $\pm$ 1.44  & 1.48 $\pm$ 0.61 & 800  \\
X~Per & 5.37 $\pm$ 0.75 & 11.0 $\pm$ 3.0$^{{\rm e}}$ & 2.74 $\pm$ 0.84 & 1000
\\
$\lambda$~Ori & 6.03 $\pm$ 2.87 & 0.13 $\pm$ 0.6 & 0.63 $\pm$ 0.42 & $\ldots$
\\   
$\chi$~Oph & 17.0 $\pm$ 3.6 & 4.27 $\pm$ 1.82 & 2.55 $\pm$ 1.21 & 400 \\

$\zeta$~Oph & 4.90 $\pm$ 1.14 & 4.47 $\pm$ 1.02 & 1.38 $\pm$ 0.45 & 400 
\\

20~Aql$^{{\rm f}}$ & 17.0 $\pm$ 2.6 & 6.55 $\pm$ 0.98 & 3.01 $\pm$ 0.64
& 850$^{{\rm g}}$ \\

\enddata
\footnotesize{
\tablenotetext{a}{Bohlin et al. (1978); Diplas \& Savage (1994).} 
\tablenotetext{b}{Savage et al. (1977).}
\tablenotetext{c}{$N_{tot}$(H) = $N$(H I) + 2 $N$(H$_2$).}
\tablenotetext{d}{Federman et al. (1994).}
\tablenotetext{e}{Mason et al. (1976).}
\tablenotetext{f}{Hanson et al. (1992); assumes 15\% error.}
\tablenotetext{g}{Knauth et al. (2001).}}
\end{deluxetable}
\clearpage

\newpage
\begin{deluxetable}{ccccc}
\tablecolumns{5}
\tablewidth{0pt}
\tablecaption{Li and K Abundances and Depletion \label{depletion}}
\tablehead{ 
\colhead{Star} & \colhead{Li/H} & \colhead{$D$(Li)} & \colhead{K/H} &
\colhead{$D$(K)}  \\
\colhead{} & \colhead{} & \colhead{[dex]} & \colhead{} & \colhead{[dex]} }
\startdata
$o$~Per & (2.8 $\pm$ 1.3) $\times$ 10$^{-10}$ & -0.9 & (2.0 $\pm$ 0.6)
$\times$ 10$^{-8}$ & -0.8 \\  
\\
X Per & (6.3 $\pm$ 2.5) $\times$ 10$^{-11}$ & -1.5 & (3.3 $\pm$ 1.0)
$\times$ 10$^{-9}$ & -1.6 \\  
\\
$\chi$ Oph & (2.4 $\pm$ 1.3) $\times$ 10$^{-10}$ & -0.9 & (1.7 $\pm$ 0.8)
$\times$ 10$^{-8}$ & -0.9 \\ 
\\
$\zeta$ Oph & (4.0 $\pm$ 1.5) $\times$ 10$^{-10}$ & -0.7 & (2.6 $\pm$
0.9) $\times$ 10$^{-8}$ & -0.7 \\ 
\\
20 Aql & (1.5 $\pm$ 0.3) $\times$ 10$^{-10}$ & -1.2 & (1.7 $\pm$ 0.4)
$\times$ 10$^{-8}$ & -0.9\\
\enddata
\end{deluxetable}
\clearpage

\newpage
\begin{deluxetable}{cccccc}
\tablecolumns{6}
\tablewidth{0pt}
\tablecaption{K/Li Abundance Ratios\label{K/Li}}
\tablehead{ 
\colhead{Star} & \colhead{\# of} & \colhead{$V_{LSR}$} &
\colhead{$N$(K)$^{{\rm a}}$} & \colhead{ $N$($^7$Li + 
$^6$Li)$^{{\rm a}}$} & \colhead{$A$(K)/$A$(Li)}  \\
\colhead{} & \colhead{components} & \colhead{[km s$^{-1}$]} & \colhead{[10$^{11}$ cm$^{-2}$]} & \colhead{[10$^8$ cm$^{-2}$]} & \colhead{} }
\startdata
$o$~Per & {\bf 2} & {\bf 3.96} & {\bf 1.46 $\pm$ 0.35} & 
{\bf 6.90 $\pm$ 2.0} & {\bf 58.8 $\pm$ 22.1} \\
 & & {\bf 6.98} & {\bf 8.96 $\pm$ 0.35} & {\bf 34.4 $\pm$ 2.1} & 
{\bf 72.4 $\pm$ 5.2} \\
  & 3 & 3.96 & 1.47 $\pm$ 0.35 & 10.8 $\pm$ 2.0 & 38.0 $\pm$ 11.5 \\
 & & 6.70 & 4.41 $\pm$ 0.35 & 12.1 $\pm$ 2.0 & 101.7 $\pm$ 18.6 \\
  & & 8.00 & 2.09 $\pm$ 0.35 & 11.2 $\pm$ 2.3 & 52.1 $\pm$ 13.8 \\
  & 4 & 4.10 & 1.47 $\pm$ 0.35 & 10.7 $\pm$ 2.0 & 38.3 $\pm$ 11.6 \\
 & & 6.83 & 4.41 $\pm$ 0.35 & 12.7 $\pm$ 2.0 & 97.2 $\pm$ 17.2 \\
 & & 8.12 & 2.09 $\pm$ 0.35 & 7.6 $\pm$ 2.0 & 76.3 $\pm$ 23.7 \\
 & & 9.10 & 0.39 $\pm$ 0.08$^{{\rm b,~c}}$ & 5.9 $\pm$ 2.0 & 18.5 $\pm$ 7.3 \\
\\
X Per & 1 & 5.85 & 30.7 $\pm$ 2.7 & 44.2 $\pm$ 3.1 & 195.0 $\pm$ 21.9 \\
  & {\bf 2} & {\bf 5.45} & {\bf 7.37 $\pm$ 0.22$^{{\rm d}}$} & {\bf 37.1 $\pm$
3.1} & {\bf 55.8 $\pm$ 5.0} \\
 & & {\bf 7.85} & {\bf 3.53 $\pm$ 0.12$^{{\rm d}}$} & {\bf 21.0 $\pm$ 3.1} &
{\bf 47.2 $\pm$ 7.2} \\
 & 4 & 4.69 & 1.70 $\pm$ 0.34$^{{\rm c,~d}}$ & 10.5 $\pm$ 3.1 & 45.5 $\pm$ 16.2 \\
 & & 5.60 & 4.60 $\pm$ 0.92$^{{\rm c,~d}}$ & 20.1 $\pm$ 3.1 & 64.2 $\pm$ 16.2 \\
 & & 7.47 & 4.90 $\pm$ 0.98$^{{\rm c,~d}}$ & 12.9 $\pm$ 3.1 & 106.6 $\pm$ 33.3 \\
 & & 8.49 & 0.18 $\pm$ 0.04$^{{\rm c,~d}}$ & 7.3 $\pm$ 3.1 & 69.2 $\pm$ 32.5 \\
\\
$\chi$ Oph & {\bf 1} & {\bf -0.30} & {\bf 12.7 $\pm$ 1.2} & {\bf 49.8 $\pm$
7.7} & {\bf 70.7 $\pm$ 12.8} \\
\\
$\zeta$ Oph & {\bf 2} & {\bf -0.62} & {\bf 5.36 $\pm$ 0.35} & 
{\bf 22.8 $\pm$ 1.0} & {\bf 66.0 $\pm$ 5.2} \\
 & & {\bf 0.42} & {\bf 2.89 $\pm$ 0.34} & {\bf 12.4 $\pm$ 1.0} & 
{\bf 65.7 $\pm$ 9.4} \\
  & 3 & -0.62 & 4.09 $\pm$ 0.82$^{{\rm b,~c}}$ & 19.2 $\pm$ 1.1 & 59.8 $\pm$
12.5 \\
 & & 0.42 & 2.72 $\pm$ 0.54$^{{\rm b,~c}}$ & 10.5 $\pm$ 1.0 & 72.8 $\pm$ 16.0
\\
  & & 1.65 & 0.11 $\pm$ 0.02$^{{\rm b,~c}}$ & $\leq$ 0.1 & $\geq$ 237.5 \\
\\
20 Aql & {\bf 2} & {\bf 1.60} & {\bf 15.6 $\pm$ 0.5} & {\bf 37.6 $\pm$ 2.1} &
{\bf 115.8 $\pm$ 7.5} \\
 & & {\bf 3.20} & {\bf 6.40 $\pm$ 0.47} & {\bf 15.1 $\pm$ 2.1} & {\bf 118.3
$\pm$ 18.6} \\
\enddata
\footnotesize{
\tablenotetext{a}{The error is 1-$\sigma$ observational uncertainty.} 
\tablenotetext{b}{Welty \& Hobbs (2001).}
\tablenotetext{c}{Assumes a 20\% uncertainty in measurement.}
\tablenotetext{d}{2 comp. S.R. Federman 2001, private communication; 
4 comp. D. Welty 2001, private communication.}}
\end{deluxetable}

\clearpage

\newpage
\begin{deluxetable}{cccc}
\tablecolumns{4}
\tablewidth{0pt}
\tablecaption{Molecular Column Densities\label{CHCNC2}}
\tablehead{ 
\colhead{Star} & \colhead{$N$(CH)} & \colhead{$N$(C$_2$)} & \colhead{$N$(CN)} \\
\colhead{} & \colhead{[cm$^{-2}$]} & \colhead{[cm$^{-2}$]} &
\colhead{[cm$^{-2}$]}}
\startdata
$o$~Per$^{~{\rm a}}$ & 1.3 $\times$
10$^{13}$ & 2.7 $\times$ 10$^{13}$ & 2.6 $\times$ 10$^{12}$   \\

40~Per$^{~{\rm b}}$ & 1.2 $\times$
10$^{13}$ &  3.6 $\times$ 10$^{12}$ & 6.4 $\times$ 10$^{11}$   \\

X~Per$^{{\rm a}}$ & 3.1 $\times$
10$^{13}$ & 5.3 $\times$ 10$^{13}$ & 8.4 $\times$ 10$^{12}$   \\

$\lambda$ Ori &  $\ldots$ &  $\ldots$ &  $\ldots$  \\

$\chi$~Oph$^{{\rm a}}$ & 3.4 $\times$ 
10$^{13}$ & 3.5 $\times$ 10$^{13}$ & 1.3 $\times$ 10$^{12}$   \\

$\zeta$~Oph$^{{\rm a}}$ & 2.5 $\times$ 
10$^{13}$ & 1.79 $\times$ 10$^{13~{\rm c}}$ & 2.6 $\times$ 10$^{12}$   \\

20~Aql$^{{\rm d}}$ & 2.0 $\times$
10$^{13}$ & 5.2 $\times$ 10$^{13}$ & 4.2 $\times$ 10$^{12}$   \\
\enddata
\footnotesize{
\tablenotetext{a}{Federman et al. (1994).} 
\tablenotetext{b}{B.-G. Andersson (2000), private communication.}
\tablenotetext{c}{Lambert, Sheffer, \& Federman (1995).}
\tablenotetext{d}{Knauth et al. (2001).}}
\end{deluxetable}
\clearpage

\newpage
\begin{deluxetable}{cccc}
\tablecolumns{4}
\tablewidth{0pt}
\tablecaption{Kinetic Temperature and Turbulent Velocity\label{Temp}}
\tablehead{ 
\colhead{Star} & \colhead{$T_k^{{\rm a}}$} & \colhead{$v_{turb}$$^{{\rm b}}$} & \colhead{$v_{turb}$$^{{\rm c}}$} \\
\colhead{} & \colhead{[K]} & \colhead{[km s$^{-1}$]} & \colhead{[km s$^{-1}$]}}
\startdata
$o$~Per & 98 $\pm$ 42 & 0.16 $\pm$ 0.10  & 0.15 $\pm$ 0.06  \\
 & 900$^{{\rm d}}$ & 0.70$^{{\rm d}}$ & 1.20$^{{\rm d}}$ \\
\\

X Per$^{{\rm d}}$ & 520 & 0.77 & 1.05 \\ 
 & 788 & 0.43 & 1.00  \\  
\\
 
$\chi$ Oph$^{{\rm d}}$ & 220 & 0.37 & 0.53 \\ 
\\
 
$\zeta$ Oph & 61 $\pm$ 27  & 0.37 $\pm$ 0.23 & 0.30 $\pm$ 0.13 \\ 
 & 63 $\pm$ 27 & 0.30 $\pm$ 0.18 & 0.21 $\pm$ 0.09 \\  
\\
 
20 Aql & 287 $\pm$ 124 & 0.47 $\pm$ 0.29 & 0.67 $\pm$ 0.29  \\
  & 770 $\pm$ 330 & 0.39 $\pm$ 0.24 & 0.97 $\pm$ 0.42 \\
\enddata
\footnotesize{
\tablenotetext{a}{$T_k$ derived from both $b$-values.} 
\tablenotetext{b}{Uses calculated $T_k$.}
\tablenotetext{c}{Assumes $T_k$ = 100 K.}
\tablenotetext{d}{Extreme values of the $b$-values ($\pm$ 1-$\sigma$) were used
to arrive at a solution for $T_k$ and $v_{turb}$; no error bars are given.}}
\end{deluxetable}
\clearpage

\newpage
\begin{deluxetable}{cccc}
\tablecolumns{4}
\tablewidth{0pt}
\tablecaption{Constraints on the Stellar $^7$Li Source$^{{\rm a}}$\label{*source}}
\tablehead{ 
\colhead{Star} & \colhead{($^7$Li/H)$_{ISM}$} & \colhead{ ($^7$Li/H)$_{GCR}$} & \colhead{($^7$Li/H)$_{stars}$}}
\startdata
$o$~Per & (1.8 $\pm$ 0.8) $\times$ 10$^{-9}$ &  (4.8 $\pm$ 3.3) $\times$
10$^{-10}$ & (1.2 $\pm$ 0.9) $\times$ 10$^{-9}$ \\
\\
X Per & (1.6 $\pm$ 0.8) $\times$ 10$^{-9}$ &  $\leq$ 6.7 $\times$
10$^{-10}$ & (8.4 $\pm$ 4.8) $\times$ 10$^{-10}$ \\   
\\ 
$\chi$ Oph & (1.8 $\pm$ 0.9) $\times$ 10$^{-9}$ & (3.8 $\pm$ 3.5)
$\times$ 10$^{-10}$ & (1.3 $\pm$ 1.4) $\times$ 10$^{-9}$  \\  
\\ 
$\zeta$ Oph & (1.9 $\pm$ 0.8) $\times$ 10$^{-9}$ & $\leq$ 2.6 $\times$
10$^{-10}$ & (1.5 $\pm$ 0.8) $\times$ 10$^{-9}$ \\  
\\
20 Aql & (1.8 $\pm$ 0.5) $\times$ 10$^{-9}$ &  $\leq$ 3.7 $\times$
10$^{-10}$ & (1.3 $\pm$ 0.4) $\times$ 10$^{-9}$ \\
\enddata
\footnotesize{
\tablenotetext{a}{Li/H abundance corrected for depletion using our depletion 
index.}}
\end{deluxetable}

\end{document}